\documentclass[11pt,fleqn]{article}
\usepackage{amsmath,amssymb,graphicx,epsfig}

\def\bea{\begin{eqnarray}}
\def\eea{\end{eqnarray}}
\def\be{\begin{equation}}
\def\ee{\end{equation}}
\def\ba{\begin{array}}
\def\ea{\end{array}}
\def\nn{\nonumber}
\def\a{& \hspace{-10pt}}
\def\b{& \hspace{-7pt}}

\def\sq{\!\raisebox{2.7pt}{\tiny $\sqrt{\;\,}\!\!\!$}}
\def\s#1{\text{\small $#1$}}
\def\t#1{\text{\tiny $#1$}}

\font\tenrsfs=rsfs10
\font\sevenrsfs=rsfs7
\font\fiversfs=rsfs5
\newfam\rsfsfam
\textfont\rsfsfam=\tenrsfs
\scriptfont\rsfsfam=\sevenrsfs
\scriptscriptfont\rsfsfam=\fiversfs
\def\mathscr#1{{\fam\rsfsfam\relax#1}}

\begin{document}

\thispagestyle{empty}

\begin{center}

$\;$

\vspace{1cm}

{\huge \bf Metastable supersymmetry breaking \\[2mm] in N=2
non-linear sigma-models}

\vspace{1.3cm}

{\Large {\bf Jean-Claude Jacot} and {\bf Claudio~A.~Scrucca}}\\[2mm] 

\vspace{0.4cm}

{\large \em Institut de Th\'eorie des Ph\'enom\`enes Physiques\\ 
Ecole Polytechnique F\'ed\'erale de Lausanne\\ 
CH-1015 Lausanne, Switzerland\\}

\vspace{0.2cm}

\end{center}

\vspace{1cm}

\centerline{\bf \large Abstract}
\begin{quote}

We perform a general study of the issue of metastability for 
supersymmetry-breaking vacua in theories with $N=1$ and $N=2$ 
global supersymmetry. This problem turns out to capture all the 
important qualitative features of the corresponding question in 
theories with local supersymmetry, where gravitational effects 
induce only quantitative modifications. Moreover, it allows to 
directly compare the conditions arising in the $N=1$ and $N=2$ 
cases, since the latter becomes particular case of the former in 
the rigid limit. Our strategy consists 
in a systematic investigation of the danger of instability coming 
from the sGoldstini scalars, whose masses are entirely due to 
supersymmetry breaking mass-splitting effects. We start by reviewing 
the metastability conditions arising in general $N=1$ non-linear 
sigma-models with chiral and vector multiplets. We then turn to 
the case of general $N=2$ non-linear sigma-models with hyper and 
vector multiplets. We first reproduce and clarify the known no-go 
theorems applying to theories with only Abelian vector multiplets 
and only hyper multiplets, and then derive new results applying 
to more general cases. To make the comparison with $N=1$ 
models as clear as possible, we rely on a formulation of $N=2$
models where one of the supersymmetries is manifestly realized 
in terms of ordinary superfields, whereas the other is realized 
through non-trivial transformations. We give a self-contained 
account of such a construction of $N=2$ 
theories in $N=1$ superspace, generalizing previous work
on various aspects to reach a general and coordinate-covariant 
construction. We also present a direct computation of the supertrace
of the mass matrix. 
\vspace{5pt}
\end{quote}

\renewcommand{\theequation}{\thesection.\arabic{equation}}

\newpage

\setcounter{page}{1}

\section{Introduction}
\setcounter{equation}{0}

One of the main issues in supersymmetric theories aiming at describing 
real fundamental interactions is how supersymmetry is spontaneously 
broken. Indeed, this breaking induces mass splittings between ordinary 
particles and their superpartners, and the details of this process are thus of 
crucial importance. It turns out that the structure of theses splittings is strongly 
constrained, and this causes some difficulties in phenomenological applications.
The perhaps most spectacular incarnation of this phenomenon is provided 
by the supertrace sum rule, which concerns the average of all the mass splittings. 
This implies for instance that renormalizable and 
anomaly-free supersymmetric extensions of the standard model cannot 
directly accommodate a viable way of spontaneously breaking supersymmetry.
The standard way out to this problem is to assume that supersymmetry 
is broken in a hidden sector, which communicates with the visible sector 
only in a way that is suppressed by some mass scale. Spontaneous 
supersymmetry breaking can then be designed in a much more flexible 
way within the hidden sector, while supersymmetry breaking effects 
communicated to the visible sector are encoded in soft supersymmetry 
breaking terms. The only strong constraints on supersymmetry breaking that one
is left with are then the metastability of the vacuum and the value of the 
cosmological constant. There are then more phenomenological constraints 
concerning the mediation of supersymmetry breaking to the visible sector 
and the structure of the soft terms. 

The problem of understanding under which conditions vacua that break 
spontaneously supersymmetry may by at least metastable clearly emerges 
as one of the most relevant possible discrimination tools on the structure 
of the hidden sector. While the stability of supersymmetry-preserving vacua
is guaranteed, that of supersymmetry-breaking ones is not, and whether they
can be metastable or even absolutely stable depends on certain particular aspects
of the theory. By now it has been well appreciated that requiring only 
metastability, rather than absolute stability, is perfectly satisfactory, as long 
as the life-time of the vacuum is sufficiently large, say larger than the age of the 
universe. Moreover, a supersymmetric theory generically admits both stable
supersymmetry-preserving vacua and metastable supersymmetry-breaking 
vacua, but generically no absolutely stable supersymmetry-breaking vacua, 
unless some extra features are imposed, like for instance the existence of a global 
$R$-symmetry \cite{NS}. This clearly means that metastability of supersymmetry-breaking 
vacua is the relevant minimal requirement to impose, rather than absolute stability. 
More specifically, the requirement of metastability translates into the requirement 
that the mass matrix of the scalar field fluctuations, given by the Hessian matrix of 
the scalar potential at the stationary point defining the vacuum, should be positive definite.
This obviously constrains the theory, but at first sight in a rather indirect and mild 
way. It turns however out that one can deduce from this requirement a quite 
simple and sharp necessary condition. 

The main observation that allows to translate the condition of metastability into 
an interesting information is the following. To a large extent, one can adjust 
the overall masses of the particles belonging to each multiplet independently
of the splittings induced by the process of spontaneous supersymmetry breaking, 
by tuning those parameters of the theory that are unrelated to the latter process.
This allows to make the square mass of most of the scalar fields arbitrarily large 
and positive. There is however one exception to this fact, represented by the 
Goldstino would-be multiplet. Indeed, for that multiplet there is an obstruction against 
changing the overall mass, due to Goldstone's theorem applied to the 
spontaneous breaking of supersymmetry. In rigid supersymmetry, this implies 
that the Goldstino is strictly massless, and the masses of its scalar partners,
the sGoldstini, are thus entirely controlled by the mass-splitting effects due to 
supersymmetry breaking. In local supersymmetry, the Goldstino is absorbed 
by the gravitino through a super-Higgs mechanism, but it remains true 
that the masses of the sGoldstini are determined by the process of supersymmetry 
breaking. This means that the only scalar fields for which there may be a potential
obstruction against achieving a positive square mass are the sGoldstini. If there
are several supersymmetries, there are just several Goldstini and thus also a larger
number of sGoldstini to look at.

The above strategy was first developed and applied to $N=1$ supergravity 
theories with only chiral multiplets in \cite{GRS1,GRS2}.\footnote{See also 
\cite{DD} for an analysis of similar spirit applied to the ideas of distribution 
and landscape of vacua.} The main outcome is that the average mass of the 
two real sGoldstini is controlled by the holomorphic sectional curvature 
of the K\"ahler manifold spanned by the scalar fields along the complex Goldstino 
direction. To achieve metastability, one then needs first of all that the scalar geometry 
admits directions along which the curvature is sufficiently small, and then that the 
Goldstino direction be sufficiently aligned towards those preferred directions. On the 
other hand, the adjustment of the value of the cosmological constant constrains the 
length of the Goldstino direction. Subsequently, this analysis was extended in 
\cite{GRS3} to more general $N=1$ theories involving both chiral and vector multiplets. 
The main conclusion is that gaugings by vector multiplets improve the situation occurring 
for just chiral multiplets, and make the bounds on the curvature milder. These metastability
conditions have then been further elaborated and applied in \cite{CGGLPS1,CGGLPS2} for 
particular classes of $N=1$ supergravity theories emerging as low-energy effective theories 
of string models, like for instance no-scale models. It has however become clear that in the 
context of string models, an analysis based on minimal $N=1$ supersymmetry may 
fail to capture all of the potentially relevant information, due to the fact that the structure 
of the low-energy effective supergravity theories underlying these models is strongly 
constrained by its higher-dimensional origin. More specifically, although for interesting 
models one gets a theory with minimal supersymmetry in four dimensions, the moduli 
sector emerging through the compactification of the extra space-time dimensions, which 
is the most natural candidate to represent the hidden sector, actually displays many of 
the features of theories with extended supersymmetry in four dimensions. As a first step 
towards gaining an understanding of the impact on the metastability condition of such 
additional peculiarities in $N=1$ theories, one may then try to study the question of 
metastability in $N=2$ theories. The case of $N=2$ supergravity theories with only 
hyper multiplets was studied in \cite{GLS}. The result of this analysis is that out of the 
four sGoldstini arising in this case, one is absorbed by the graviphoton and is thus 
not dangerous, whereas the other three have an average square mass which is 
negative when the cosmological constant is positive, meaning that it is impossible
to achieve metastability. A similar no-go theorem has been known for a long time to arise
also in $N=2$ theories involving only Abelian vector multiplets \cite{Many}. On the other
hand, it has been shown through the construction of particular examples that more
general $N=2$ theories involving non-Abelian vector multiplets and/or both hyper 
and vector multiplets can admit metastable supersymmetry-breaking vacua with
positive cosmological constant \cite{FTV,O}. A natural step to take is then to try to 
understand the metastability condition applying for general $N=2$ theories, with the 
aim of figuring out which are the truly necessary ingredients to go in business. Such 
an analysis is however quite challenging from a technical point of view \cite{DGLS}.

The aim of this work is to study the question of metastability in theories with 
$N=1$ and $N=2$ global supersymmetry. This rigid limit of the problem turns 
out to capture all the qualitatively important aspects of the corresponding 
problem in local supersymmetry, gravitational effects being responsible 
only for a quantitative deformation of the results. Moreover, besides yielding
a much simpler and more transparent setting, the rigid limit also offers the 
very interesting possibility of directly comparing the results for $N=2$ theories 
to those of $N=1$ theories. This is due to the fact that in global supersymmetry
$N=2$ theories with hyper and vector multiplets are just particular cases of
$N=1$ theories with chiral and vector multiplets, whereas on the contrary in local 
supersymmetry this is not the case, due to the effects of the spin-$3/2$ multiplet 
describing the degrees of freedom needed to complete the $N=1$ gravitational 
multiplet to the $N=2$ one. To perform this study and make the 
comparison between $N=2$ and $N=1$ theories as transparent as possible, we 
shall use a formulation of $N=2$ theories based on $N=1$ superspace, where 
one of the supersymmetries is manifestly realized in terms of ordinary superfields,
whereas the other is realized by a non-trivial transformation mixing different 
superfields. We will follow the approach of \cite{HKLR}, and generalize 
it in such a way to reach a general and coordinate-covariant 
construction, which in components reproduces the rigid limit of the general 
$N=2$ supergravity theory as formulated in \cite{DFF,ABCDFFM}. We will then 
use the same strategy as in previous supergravity studies and work out the 
metastability conditions by systematically computing the masses of all the scalar 
sGoldstini. We will also revisit the computation of the supertrace sum rule, since it 
represents a related information, and rederive in a more direct way the results that 
were indirectly deduced in \cite{GRK} from a superspace evaluation of the quadratic 
divergence in the one-loop effective action. We shall use the conventions of \cite{WB}.

The paper is organized as follows. In section 2 we present the simplest case 
of $N=1$ theories with only chiral multiplets and discuss the rigid version of 
the results of \cite{GRS1,GRS2}. In section 3 we present the case of $N=1$  
theories with both chiral and vector multiplets, and describe the rigid version 
of the result of \cite{GRS3}, generalized to non-Abelian gauge groups. 
In section 4 we consider the case of $N=2$ 
theories with only hyper multiplets, formulated as particular cases of $N=1$ 
theories with only chiral multiplets. We derive the analogue of the result
of \cite{GLS} in the rigid limit and clarify how its emerges when gravity is 
decoupled and which information is associated respectively to the minimal 
and to the additional supersymmetries. In section 5 
we consider the case of $N=2$ theories with only vector multiplets, 
formulated as particular cases of $N=1$ theories with chiral and vector 
multiplets. After recovering the rigid limit of the result of \cite{Many} in the 
Abelian case, we study the non-Abelian case and derive a new result 
applying to this situation, discussing again carefully which information comes 
from the minimal supersymmetry and which from the additional one. 
In section 6, we finally consider the case of general 
$N=2$ theories with both hyper and vector multiplets. We set up the logic 
of the study of the metastability condition, and discuss the form that it is 
expected to take. Finally, in section 7 we present our conclusions.

\section{N=1 models with chiral multiplets}
\setcounter{equation}{0}

Let us start by considering the simplest case of $N=1$ theories with $n_{\rm C}$ 
chiral multiplets $\Phi^i$. The most general two-derivative Lagrangian is specified 
in terms of a real K\"ahler potential $K$ and a holomorphic superpotential $W$, and reads:
\bea
\mathcal{L} = \int \! d^4 \theta\, K(\Phi,\bar \Phi) + \int \! d^2 \theta\, W(\Phi) + {\rm h.c.} \,.
\eea
In components, this gives
\bea
\mathcal{L} \b=\b-g_{i \bar \jmath} \, \partial_\mu \phi^i \partial^\mu \bar \phi^{\bar \jmath}
-ig_{i\bar \jmath} \, \psi^i \big(\partial\!\!\!/\bar \psi^{\bar \jmath} + 
\Gamma^{\bar \jmath}_{\bar m \bar n} \, \partial\!\!\!/ \bar \phi^{\bar m} \bar \psi^{\bar n} \big) 
- V_{\rm S} - V_{\rm F} \,,
\eea
where $g_{i \bar \jmath} = K_{i \bar \jmath}$ defines a K\"ahler 
geometry for the scalar manifold \cite{Zumino} 
and\footnote{Our conventions for the curvature are such that the non-vanishing components 
of the Riemann tensor are given by
$R_{i \bar \jmath k \bar l} = K_{i \bar \jmath k \bar l} - g^{\bar s r} K_{i k \bar s} K_{\bar \jmath \bar l r}$
and those of the Ricci tensor by $R_{i \bar \jmath} = - g^{k \bar l} R_{i \bar \jmath k \bar l}$.} 
\bea
V_{\rm S} \b=\b g^{i\bar \jmath} \, W_i \bar W_{\bar \jmath} \,, \\
V_{\rm F} \b=\b \s{\frac 12} \nabla_i W_j \, \psi^i \psi^j + {\rm h.c.}
- \s{\frac 14} R_{i \bar \jmath k \bar l} \, \psi^i \psi^k \bar \psi^{\bar \jmath} \bar \psi^{\bar l} \,,
\eea

The supersymmetry transformation laws are defined by the action of the supercharges on 
the superfields and act as follows in components:
\bea
\a\a \delta \phi^i = \s{\sqrt{2}} \, \epsilon \,\psi^i \,, \\
\a\a \delta \psi^i = \s{\sqrt{2}}\, \epsilon \,F^i + \s{\sqrt{2}}i\,\partial\!\!\!/ \phi^i \, \bar \epsilon \,. 
\eea
The auxiliary fields $F^i$ are given by
\be
F^i = - g^{i \bar \jmath} \, \bar W_{\bar \jmath } + \s{\frac 12} \Gamma^i_{jk}\, \psi^j \psi^k \,.
\ee

The extension to supergravity is well known \cite{sugra1,sugra2} and does not present 
particularly subtle features. In particular, any model of the above type can be consistently 
coupled to gravity. The main new feature is that there appears a non-trivial $U(1)$ 
bundle over the scalar manifold, whose curvature is proportional to $M_{\rm P}^{-2}$, 
and the manifold becomes K\"ahler-Hodge.

\subsection{Supertrace}

At a generic point in the scalar field space and for vanishing fermions, the auxiliary fields simplify to
\be
F^i = - \bar W^i \,.
\ee
The mass matrix of the scalar fields is given by the following two blocks:
\bea
\a\a (m_0^2)_{i \bar \jmath} = \nabla_i W_k \nabla_{\bar \jmath} \bar W^k - R_{i \bar \jmath k \bar l} \, F^k \bar F^{\bar l} \,, \\
\a\a (m_0^2)_{ij} = - \nabla_i \nabla_j W_k \, F^k + \Gamma_{ij}^k V_{{\rm S}k}\,.
\eea
The mass matrix of the fermions is instead
\bea
\a\a (m_{1/2})_{ij} = \nabla_i W_j \,.
\eea
One easily computes
\bea
\a\a {\rm tr} [m_0^2] = 2\, \nabla_i W_j \nabla^i \bar W^j + 2\, R_{i \bar \jmath} \, F^i \bar F^{\bar \jmath} \,,\\
\a\a {\rm tr} [m_{1/2}^2] = \nabla_i W_j \nabla^i \bar W^j \,.
\eea
It follows that the supertrace of the mass matrix is given by \cite{GRK}
\bea
{\rm str}[m^2] \b\equiv\b {\rm tr} [m_0^2] - 2\, {\rm tr} [m_{1/2}^2] \nn \\
\b=\b 2\, R_{i \bar \jmath} \, F^i F^{\bar \jmath} \,.
\label{strc}
\eea

\subsection{Metastability}

The possible vacua of the theory correspond to points in the scalar manifold that 
satisfy the stationarity condition $V_{{\rm S}i} = 0$, which reads:
\be
\nabla_i W_j \,F^j = 0 \,.
\ee

On the vacuum $\delta \psi^i = \s{\sqrt{2}} \epsilon F^i$, and supersymmetry is 
spontaneously broken if some of the auxiliary fields $F^i$ are non-vanishing.
The order parameter is the norm of the vector of auxiliary fields, which defines 
the scalar potential energy $V_{\rm S} = F^i \bar F_i$. In such a situation, there 
is then a massless Goldstino fermion given by:
\be
\eta = \s{\sqrt{2}}\, \bar F_i \psi^i \,.
\ee
Indeed, the stationarity condition directly implies that this is a flat direction of the fermion 
mass matrix:
\be
m_\eta = 0 \,.
\ee
The two would-be supersymmetric scalar partners of this fermionic mode, 
the sGoldstini, generically have non-zero masses, but these are controlled 
by the process of supersymmetry breaking, and cannot be affected by 
supersymmetric mass terms in the superpotential. These modes are then 
particularly dangerous for the metastability of the vacuum. From the form 
of the supersymmetry transformations, we see that they can be 
parametrized by the two independent real linear combinations that one 
can form with the complex Goldstino vector $\eta^i = \sqrt{2} F^i$, namely:
\be
\varphi_+ = \bar F_i \phi^i + F_{\bar \imath} \bar \phi^{\bar \imath} \,,\;\;
\varphi_- = i \bar F_i \phi^i - i F_{\bar \imath} \bar \phi^{\bar \imath} \,.
\ee
The masses of these two scalar modes can now be computed by evaluating 
the scalar mass matrix along the directions $\varphi_+^I = (F^i, \bar F^{\bar \imath})$ 
and $\varphi_-^I = (i F^i, - i \bar F^{\bar \imath})$, and dividing by the length of these 
vectors, which is $2 F^i \bar F_i$. After using the stationarity condition to simplify the 
results, one obtains:
\be
m_{\varphi_\pm}^2 = R \,F^i \bar F_i \pm \Delta \,. \label{m+-F}
\ee
The first term involving the quantity $R$ comes from the contribution of the Hermitian block 
$(m_0^2)_{i \bar \jmath}$ of the mass matrix, and it turns out that $R$ is simply the holomorphic 
sectional curvature of the scalar manifold in the complex plane defined by the Goldstino direction 
$F^i$ of supersymmetry breaking:
\be
R = - \frac {R_{i \bar \jmath m \bar n} F^i \bar F^{\bar \jmath} F^m \bar F^{\bar n}}{(F^k \bar F_k)^2} \,.
\label{RF}
\ee
The second term $\Delta$ corresponds instead to the contribution from the complex block 
$(m_0^2)_{i j}$, and has a more complicated expression, which depends also on second 
and third derivatives of the superpotential and is thus much more model-dependent.
But happily, we see that the average of the two masses is independent of $\Delta$, and 
one thus finds the following result, which defines an upper bound on the lowest mass
eigenvalue:
\be
m^2_{\varphi} \equiv \s{\frac 12} \big(m_{\varphi_+}^2 \!+ m_{\varphi_-}^2\big) = R \, F^i \bar F_i \,.
\label{metc}
\ee
From this result, we conclude that a necessary condition for not having a tachyonic 
mode is that the holomorphic sectional curvature $R$ be positive.\footnote{In the 
limiting case of models based on a flat geometry, for which $R$ vanishes, one 
generically finds that one of the sGoldstini is tachyonic and the other not. The best 
thing that one may do is then to tune the superpotential to make both of them massless,
with vanishing $\Delta$. One can then show that in such a situation the sGoldstini are 
not only massless, but actually correspond to flat directions of the potential and are 
identified with the so-called pseudo-moduli arising in these models. See \cite{Ray} for 
a recent discussion.} 

The above result is the rigid limit of the result obtained in \cite{GRS1,GRS2} for the 
supergravity case. Introducing the gravitino mass $m_{3/2}$ and the Planck mass 
$M_{\rm P}$, the cosmological constant reads $V_{\rm S} = F^i \bar F_i - 3\, m_{3/2}^2 M_{\rm P}^2$ 
and the average sGoldstino mass is given by the following formula in supergravity
\be
m^2_{\varphi} = R\,F^i \bar F_i + 2\,m_{3/2}^2 \,. 
\ee
We see that the main feature of this result, namely the dependence on the curvature 
$R$, is also captured in the rigid limit, in which $m_{3/2} \to 0$ and $M_{\rm P} \to \infty$. 
Gravitational effects influence only quantitatively the result, and the metastability 
condition implies now that the holomorphic sectional curvature $R$ be larger than 
the negative critical value $-2\, m_{3/2}^2/(V_{\rm S} + 3\, m_{3/2}^2 M_{\rm P}^2)$, which 
tends to $0$ in the rigid limit.

The above necessary condition for metastability becomes also sufficient if for a given 
K\"ahler potential $K$ one allows the superpotential $W$ to be adjusted \cite{CGGLPS1,CGGLPS2}. 
Indeed, at the stationary point one may tune $W_i$ to maximize the average sGoldstino mass, 
$W_{ij}$ to make the other masses arbitrarily large, and $W_{ijk}$ to set the splitting between 
the two sGoldstino masses to zero. 

\section{N=1 models with chiral and vector multiplets}
\setcounter{equation}{0}

Let us consider next the most general case of $N=1$ theories with $n_{\rm C}$ chiral multiplets $\Phi^i$
and $n_{\rm V}$ vector multiplets $V^{\rm a}$. The most general two-derivative Lagrangian is in this case 
specified by a real K\"ahler potential $K$, a holomorphic superpotential $W$, a holomorphic 
gauge kinetic function $f_{ab}$, some holomorphic Killing vectors $X_a^i$ and some real Fayet-Iliopoulos 
constants $\xi_a$, and reads:
\be
\mathcal{L} = \int \! d^4 \theta\, \Big[K(\Phi,\bar \Phi,V) + \xi_a V^a \Big]
+ \int \! d^2 \theta\, \Big[W(\Phi) + \s{\frac 14}\, f_{ab} (\Phi) \,W^{a \alpha} W_\alpha^b \Big] + {\rm h.c.} \,.
\ee
The gauge transformations form a Lie group with structure constants $f_{ab}^{\;\;\;c}$, and act as 
follows on the superfields, with chiral multiplet parameters $\Lambda^a$:
\bea
\a\a \delta_{\rm g} \Phi^i = \Lambda^a X_a^i(\Phi) \,, \\
\a\a \delta_{\rm g} V^a = - \frac i2 \big(\Lambda^a - \bar \Lambda^a) 
+ \frac 12 f_{bc}^{\;\;\;a} \big(\Lambda^b + \bar \Lambda^b) V^c + {\cal O}(V^2)\,.
\eea
Gauge invariance of the Lagrangian imposes that the variation of the non-holomorphic terms 
should be at most a K\"ahler transformation of the form $\Lambda^a f_a + \bar \Lambda^a \bar f_a$, 
where the $f_a$ are some holomorphic functions, whereas the holomorphic terms should be 
strictly invariant. This implies the following conditions:
\bea
\a\a X_a^i K_i - \s{\frac i2} K_a = f_a \,, \\[-0.5mm]
\a\a X_a^i W_i = 0 \,, \label{Winv} \\[0.5mm]
\a\a X_a^i f_{bci} = - 2 f_{a\t{(}b}^{\;\;\;\;d} f_{c\t{)}d} \,, \\[0.5mm]
\a\a \xi_a = 0 \; \mbox{whenever} \; f_{bc}^{\;\;\;a} \neq 0 \,.
\eea
These equations show that $-\frac 12 K_a$ can be identified with the real Killing potential
for the Killing vector $X_a^i$, and the Fayet-Iliopoulos constants $\xi_a$ can be interpreted 
as coming from the freedom of adding a constant to this potential for Abelian generators:
\be
X_a^i = \s{\frac i2}\, g^{i \bar \jmath} \nabla_{\bar \jmath} K_a\,,\;\;
\ee
One also has to impose the equivariance condition on the Killing vectors, i.e. that 
the operators $\delta_a = X_a^i \partial_i + \bar X_a^{\bar \jmath} \partial_{\bar \jmath}$
satisfy the group algebra $[\delta_a,\delta_b] = - f_{ab}^{\;\;\;c} \delta_c$. This guarantees 
that the Killing potentials can be chosen to transform in the adjoint representation, so that
\be
g_{i \bar \jmath} X_{\t{[}a}^i \bar X_{b\t{]}}^{\bar \jmath} = \s{\frac i4}\, f_{ab}^{\;\;\; c} K_c 
\label{equiv}\,.
\ee
In the Wess-Zumino gauge, the action simplifies to the following expression:
\bea
\mathcal{L} \b=\b \int \! d^4 \theta\, \Big[K(\Phi,\bar \Phi) + \big(K_a (\Phi,\bar \Phi) + \xi_a \big) V^a 
+ 2\, g_{i \bar \jmath} (\Phi,\bar \Phi) X_a^i (\Phi) \bar X_b^{\bar \jmath} (\bar \Phi) V^a V^b \Big] \nn \\
\b\;\b + \int \! d^2 \theta\, \Big[W(\Phi) + \s{\frac 14}\, f_{ab} (\Phi) \,W^{a \alpha} W_\alpha^b \Big] + {\rm h.c.} \,.
\eea
In components, this gives
\bea
\mathcal{L} \b=\b - g_{i \bar \jmath} \, D_\mu \phi^i D^\mu \bar \phi^{\bar \jmath}
- \s{\frac 14} h_{ab} \, F^a_{\mu\nu} F^{b\mu\nu} + \s{\frac 14} k_{ab} \, F^a_{\mu\nu}\tilde{F}^{b\mu\nu} 
-ig_{i \bar \jmath} \, \psi^i \big(D\!\!\!\!/\,\bar \psi^{\bar \jmath} + 
\Gamma^{\bar \jmath}_{\bar m \bar n} \, D\!\!\!\!/\, \bar \phi^{\bar m} \bar \psi^{\bar n} \big) 
\nn \\
\b\;\b -\frac i2 h_{ab} \, \lambda^{a} D\!\!\!\!/\, \bar \lambda^b + {\rm h.c.} 
+ \s{\frac 1{\!\sqrt{2}}} h_{abi} \, \lambda^a \sigma^{\mu \nu} \psi^i F^{b}_{\mu\nu} + {\rm h.c.} 
- V_{\rm S} - V_{\rm F}\,, 
\eea
where:
\bea
V_{\rm S} \b=\b g^{i \bar \jmath} \, W_i \bar W_{\bar \jmath} + \s{\frac 18} \, h^{ab}(K_a \!+ \xi_a)(K_b \!+ \xi_b) \nn \\[-0.5mm]
V_{\rm F} \b=\b \s{\frac 12} \Big[\nabla_i W_j \, \psi^i \psi^j - g^{i \bar \jmath} h_{abi} \bar W_{\bar \jmath} \, \lambda^a \lambda^b
+ \s{\sqrt{8}} \big(g_{i \bar \jmath} \bar X_a^{\bar \jmath} + \s{\frac i4} h^{bc} h_{abi} (K_c \!+ \xi_c) \big) \psi^i \lambda^a \Big] + {\rm h.c.} \nn \\
\b\;\b - \s{\frac 14} R_{i \bar \jmath k \bar l} \, \psi^i \psi^k \bar \psi^{\bar \jmath} \bar \psi^{\bar l}
+ \s{\frac 14} g^{i \bar \jmath} h_{abi} h_{cd \bar \jmath} \, \lambda^a \lambda^b \bar \lambda^c \bar \lambda^d 
+ \s{\frac 12} h^{cd} h_{aci} h_{bd \bar \jmath} \, \psi^i \lambda^a \bar \psi^{\bar \jmath} \bar \lambda^b \nn \\
\b\;\b - \s{\frac 14} \Big[\nabla_i h_{abj} \, \psi^i \psi^j \lambda^a \lambda^b + 
h^{cd} h_{aci} h_{bdj} \psi^i \lambda^a \psi^j \lambda^b \Big] + {\rm h.c.} \,.
\eea
In these formulae, $D_\mu$ is the gauge covariant derivative acting as 
$D_\mu \phi^i = \partial_\mu \phi^i + A^a_\mu X_a^i$, $D_\mu \psi^i = \partial_\mu \psi^i + A^a_\mu \partial_j X_a^i \,\psi^j$
and $D_\mu \lambda^a = \partial_\mu \lambda^a + f_{bc}^{\;\;\;a} A_\mu^b \lambda^c$, 
$F^a_{\mu \nu} = \partial_\mu A^a_\nu - \partial_\nu A_\mu^a + f_{bc}^{\;\;\;a} A_\mu^b A_\nu^c$ is the field-strength,
whereas $h_{ab}$ and $k_{ab}$ are the real and imaginary parts of $f_{ab}$. 

The supersymmetry transformation laws involve not only the usual action of the supercharges, but also a 
compensating gauge transformation needed to preserve the Wess-Zumino gauge choice, with superfield parameter 
given by $\Lambda^{a} = 2 i \theta \sigma^\mu \bar \epsilon A^a_\mu + 2 \theta^2 \bar \epsilon \bar \lambda^a$.
The additional gauge transformation has no effect on the transformation laws of the components of $V^a$, but gives some 
additional terms in those of the components of $\Phi^i$. In particular, it turns the ordinary derivative 
appearing in $\delta \psi^i$ into a gauge-covariant derivative. One finally finds
\bea
\a\a \delta \phi^i = \s{\sqrt{2}}\, \epsilon\, \psi^i \,, \\
\a\a \delta \psi^i = \s{\sqrt{2}}\, \epsilon\, F^i + \s{\sqrt{2}}i\,D\!\!\!\!/\, \phi^i \, \bar \epsilon \,,\\
\a\a \delta A_\mu^a = i \epsilon\, \sigma_\mu \bar \lambda^a - i \lambda^a \sigma_\mu \, \bar \epsilon \,, \\
\a\a \delta \lambda^a = i \epsilon\, D^a + \sigma^{\mu \nu} \epsilon\, F_{\mu \nu}^a \,.
\eea
The auxiliary fields $F^i$ and $D^a$ are given by
\bea
\a\a F^i = - g^{i \bar \jmath} \bar W_{\bar \jmath} + \s{\frac 12} \Gamma^i_{jk} \, \psi^j \psi^k 
+ \s{\frac 12} g^{i \bar \jmath} h_{ab \bar \jmath} \, \bar \lambda^a \bar \lambda^b \,, \\
\a\a D^a = - \s{\frac 12} h^{ab} (K_b \!+ \xi_b) - \s{\frac i{\!\sqrt{2}}} h^{ab} h_{bci}\, \psi^i \lambda^c + {\rm h.c.} \,.
\eea

The extension to supergravity is again well known \cite{sugragau1,sugragau2} and presents in this 
case a subtlety. It turns out that models of the above type can generically be coupled to gravity 
only in the absence of Fayet-Iliopoulos terms, i.e. when $\xi_a = 0$. This is due to the 
fact that the accidental gauge-invariance of this term in rigid supersymmetry is spoiled 
by gravitational effects. Similarly, there cannot be any non-trivial holomorphic function 
appearing in gauge transformations of the K\"ahler potential, once the superpotential is 
assumed to be gauge invariant, and one needs $f_a = 0$. A way out of this restriction 
arises only if the theory admits an $R$-symmetry, which can be gauged and for which 
a Fayet-Iliopoulos term is possible \cite{FI1,FI2}. For the rest, the main new feature is as before 
that there appears a non-trivial $U(1)$ bundle over the scalar manifold with curvature 
proportional to $M_{\rm P}^{-2}$, and the manifold becomes K\"ahler-Hodge. From now 
on, we shall restrict to models that can emerge from a smooth rigid limit of the local case, 
although most of the results that we shall derive in the remainder of this section have a 
more general validity. We shall moreover not discuss the special possibility of gauging a 
$U(1)_R$ symmetry, and thus require for simplicity that
\be
\xi_a = 0 \,,\;\; f_a = 0\,.
\ee

\subsection{Supertrace}

At a generic point in the scalar field space and for vanishing fermions and vector fields, the auxiliary fields simplify to
\bea
\a\a F^i = - \bar W^i \,,\\
\a\a D^a = - \s{\frac 12} h^{ab} K_b \,.
\eea
The mass matrix of the scalar fields is given by the following two blocks:
\bea
\a\a (m_0^2)_{i \bar \jmath} = \nabla_i W_k \nabla_{\bar \jmath} \bar W^k - R_{i \bar \jmath k \bar l} \, F^k \bar F^{\bar l} 
+ h^{ab} \bar X_{a i} X_{b \bar \jmath} + h^{ab} h_{aci} h_{bd \bar \jmath} \, D^b D^c \nn \\[0mm]
\a\a \hspace{45pt} +\, \s{\frac i2} \big(\nabla_i X_{a \bar \jmath} - 2 h^{bc} h_{abi} X_{c \bar \jmath} \big)\,D^a + {\rm h.c.} \,, \\[0mm]
\a\a (m_0^2)_{ij} = - \nabla_i \nabla_j W_k \, F^k - h^{ab} \bar X_{a i} \bar X_{b j} 
- \s{\frac 12} \big(\nabla_i h_{abj} - 2 h^{cd} h_{aci} h_{bdj} \big) \, D^a D^b \nn \\[0.5mm]
\a\a \hspace{45pt} +\,2 i\, h^{bc} h_{ab\t{(}i} \bar X_{cj\t{)}} D^a 
+ \Gamma_{ij}^k V_{{\rm S}k} \,.
\eea
The mass matrix of the fermions involves instead the following three blocks:
\bea
\a\a (m_{1/2})_{ij} = \nabla_i W_j \,, \\[1.5mm]
\a\a (m_{1/2})_{ab} = h_{abi} \, F^i \,, \\[0mm]
\a\a (m_{1/2})_{ia} = \s{\sqrt{2}}\, \bar X_{ai} - \s{\frac i{\!\sqrt{2}}} h_{abi} \, D^b \,.
\eea
Finally, the mass matrix of the vectors is
\bea
\a\a (m_1^2)_{ab} = 2\, X_{\t{(}a}^i \bar X_{b\t{)}i} \,.
\eea
A straightforward computation gives
\bea
\a\a {\rm tr} [m_0^2] = 2\, \nabla_i W_j \nabla^i \bar W^j + 2\, R_{i \bar \jmath} \, F^i \bar F^{\bar \jmath} 
+ 2\, h^{ab} \bar X_{a i} X_b^i + 2 h^{ab} h_{aci} h_{bd}^{\;\;\;i} \, D^b D^c \nn \\[1mm]
\a\a \hspace{42pt} +\, i \big(\nabla_i X_a^i - 2 h^{bc} h_{abi} X_c^i \big) D^a + {\rm h.c.} \,,\\[1.5mm]
\a\a {\rm tr} [m_{1/2}^2] = \nabla_i W_j \nabla^i \bar W^j 
+ h^{ac} h^{bd} h_{abi} h_{cd \bar \jmath} \, F^i \bar F^{\bar \jmath} + 4\, h^{ab} \bar X_{ai} X_b^i \nn \\[0.5mm]
\a\a \hspace{52pt} +\, h^{cd} h_{aci} h_{bd}^{\;\;\;i} \, D^a D^b
- 2 i\, h^{ab} h_{bci} X_a^i D^c + {\rm h.c.} \,, \\[1mm]
\a\a {\rm tr} [m_1^2] = 2\, h^{ab} \bar X_{ai} X_b^i \,.
\eea
It follows that the supertrace of the mass matrix is given by \cite{GRK}
\bea
{\rm str}[m^2] \b\equiv\b {\rm tr} [m_0^2] - 2\, {\rm tr} [m_{1/2}^2] + 3\, {\rm tr} [m_1^2] \nn \\[1mm]
\b=\b  2 \big(R_{i \bar \jmath} - h^{ac} h^{bd} h_{abi} h_{cd \bar \jmath} \big) F^i \bar F^{\bar \jmath}
+ i \big(\nabla_i X_a^i + 2 h^{bc} h_{abi} X_c^i \big) D^a + {\rm h.c.} \,.
\label{strcv}
\eea
Note that we did not need to fix a gauge for the ordinary gauge symmetry to perform this computation, 
thanks to the fact that the unphysical would-be Goldstone scalars that are eaten by the gauge fields 
correspond to flat directions of the scalar mass matrix. By tracing over the whole $m_0^2$, one does 
therefore not overcount these modes, since they come with a vanishing value of the mass. 

\subsection{Metastability}

The possible vacua of the theory correspond to points in the scalar manifold that 
satisfy the stationarity condition $V_{{\rm S}i} = 0$, which implies
\be
\nabla_i W_j \,F^j  + \s{\frac 12} h_{abi} D^a D^b + i \bar X_{a i} D^a = 0 \,.
\ee
By contracting this relation with the Killing vectors $X_a^i$ and taking the imaginary part,
and using (\ref{Winv}) and its derivative as well as (\ref{equiv}),
one also finds the following relation between the values of the $F^i$ and $D^a$ auxiliary fields:
\be
i \nabla_i X_{a \bar \jmath} \, F^i \bar F^{\bar \jmath} - X_{\t{(}a}^i X_{b\t{)} i} \, D^b 
+ \s{\frac 12} f_{ab}^{\;\;\;d} k_{dc} \,D^b D^c = 0 \,. \label{relDFF}
\ee
By further contraction with $D^a$, this also implies
$i \nabla_i X_{a \bar \jmath} \, D^a F^i \bar F^{\bar \jmath} - X_{\t{(}a}^i X_{b\t{)} i} \, D^a D^b = 0$.
This formula shows in particular that if the $F^i$ vanish then also the $D^a$ vanish, under the 
assumption that there are neither Fayet-Iliopoulos terms nor non-trivial K\"ahler transformation 
functions associated to gauge transformations. Indeed, in such a situation the first term vanishes, 
and the equation implies that either $D^a$ or $X_a^i$ should vanish. But $X_a^i = 0$
implies also $D_a = 0$, whenever the total non-holomorphic term in the Lagrangian is 
strictly gauge invariant, since in that case $D_a = - i X_a^i K_i$.

On the vacuum one has $\delta \psi^i = \s{\sqrt{2}} \epsilon F^i$ and $\delta \lambda^a = i \epsilon D^a$,
and supersymmetry is spontaneously broken if at least some of the auxiliary fields $F^i$ or $D^a$ are 
non-vanishing. The order parameter is the norm of the vector of auxiliary fields, which defines the 
scalar potential energy $V_{\rm S} = F^i \bar F_i + \frac 12 D^a D_a$. 
In such a situation, there is then a massless Goldstino given by
\be
\eta = \s{\sqrt{2}} \bar F_i \psi^i + i D_a \lambda^a \,.
\ee
Indeed, the stationarity condition and the gauge invariance of the superpotential 
imply that this is a flat direction of the fermion mass matrix:
\be
m_\eta = 0 \,.
\ee
As before, the would-be supersymmetric partners of this fermionic mode, the 
sGoldstini, have masses that are controlled by the process of supersymmetry breaking. 
They are then particularly dangerous for the metastability of the vacuum. From the form 
of the supersymmetry transformations, we see that in this case these modes are linear 
combinations of both scalars and vectors. However, since the vector components cannot
get negative square masses, the relevant thing to look at is the projection onto the scalar
field space. One then gets the same two independent real linear combinations as before,
corresponding to the projection of the complex Goldstino vector $\eta^i = \s{\sqrt{2}} F^i$:
\be
\varphi_+ = \bar F_i \phi^i + F_{\bar \imath} \bar \phi^{\bar \imath} \,,\;\;
\varphi_- = i \bar F_i \phi^i - i F_{\bar \imath} \bar \phi^{\bar \imath} \,.
\ee
The masses of these two scalar modes can now be computed as before, 
by evaluating the scalar mass matrix along the directions $\varphi_+^I = (F^i, \bar F^{\bar \imath})$
and $\varphi_-^I = (i F^i, - i \bar F^{\bar \imath})$, and dividing by the length of 
these vectors, which is $2 F^i \bar F_i$. After using the stationarity condition as 
well as the various constraints imposed by gauge invariance to simplify the results, 
one obtains
\be
m_{\varphi_\pm}^2 = R \,F^i \bar F_i + S\, D^a D_a + \s{\frac 14}\, T\, \frac {(D^a D_a)^2}{F^i \bar F_i} 
+ M^2 \frac {D^a D_a}{F^i \bar F_i} \pm \Delta \,. \label{m+-FD}
\ee
The first four terms involving the quantities $R$, $S$, $T$ and $M^2$ come from the contribution 
of the Hermitian block $(m_0^2)_{i \bar \jmath}$ of the mass matrix. It turns out that $R$ is as before 
the holomorphic sectional curvature in the complex plane defined by the Goldstino direction $F^i$,
whereas $S$ and $T$ and similar objects defined out of the derivatives of $h_{ab}$, and $M^2$
is related to the mass of the vector fields:
\bea
\a\a R = - \frac {R_{i \bar \jmath m \bar n} \,F^i \bar F^{\bar \jmath} F^m \bar F^{\bar n}}{(F^k \bar F_k)^2} \,, \label{RFD} \\
\a\a S = \frac {h_{aci} h^{cd} h_{db \bar \jmath} \,F^i \bar F^{\bar \jmath} D^a D^b }{(F^k \bar F_k)(D^c D_c)} \,, \label{SFD} \\
\a\a T = \frac {h_{abi} h_{cb}^{\;\;\;i} D^a D^b D^c D^d}{(D^e D_e)^2} \,, \label{TFD} \\
\a\a M^2 = \frac {2 X_a^i \bar X_{bi}\, D^a D^b}{D^c D_c} \,. \label{MFD}
\eea
The quantity $\Delta$ corresponds instead to the contribution from the complex block 
$(m_0^2)_{i j}$, and has again a more complicated and model-dependent expression.
But as before, we see that the average of the two masses is independent of $\Delta$, and 
one thus finds the following result, which defines an upper bound on the lowest 
mass eigenvalue:
\be
m^2_{\varphi} \equiv \frac 12 \big( m_{\varphi_+}^2 + m_{\varphi_-}^2\big) = R \,F^i \bar F_i + S\, D^a D_a 
+ \s{\frac 14}\,T\, \frac {(D^a D_a)^2}{F^i \bar F_i} + M^2 \frac {D^a D_a}{F^i \bar F_i}  \,.
\ee
From this, we conclude that a necessary condition for not having a tachyonic mode 
is that the holomorphic sectional curvature $R$ be larger than a certain negative-definite 
value controlled by the data of the gauge sector.

In this case, there is an additional feature concerning scalar fields that has to be considered. 
Indeed, on the vacuum one has $\delta_g \phi^i = \lambda^a X_a^i$, and some of the gauge 
symmetries may be spontaneously broken if some of the components of $X_a^i$ are 
non-vanishing. The order parameters are the eigenvalues of the matrix of scalar products 
of the Killing vectors, which defines the gauge boson mass matrix 
$(m_1^2)_{ab} = 2\,X_{\t{(}a}^i \bar X_{b\t{)}i}$. In such a situation, there are thus also other complex 
directions of special relevance, namely those defined by the Killing vectors $X_a^i$. These are 
related to the would-be Goldstone modes that are eaten by the massive vector fields when the gauge 
symmetry is spontaneously broken, which are given by the following real combinations:
\be
\sigma_a = \bar X_{ai} \phi^i + X_{a \bar \imath} \bar \phi^{\bar \imath} \,.
\ee
Along these unphysical directions, the scalar mass matrix has vanishing value:
\be
m^2_{\sigma_a} = 0 \,.
\ee
One may then wonder what happens along the conjugate directions defined by
\be
\rho_a = i \bar X_{ai} \phi^i - i X_{a \bar \imath} \bar \phi^{\bar \imath} \,.
\ee
These generically have non-vanishing masses, 
\be
m_{\rho_a}^2 \neq 0 \,.
\ee
These informations all directly follow from the gauge invariance of the scalar potential.
This implies that $X_a^i \nabla_i V_{\rm S} + \bar X_a^{\bar \jmath} \nabla_{\bar \jmath} V_{\rm S} = 0$,
and can be checked to be a consequence of gauge invariance conditions listed previously
plus the equivariance condition. Taking then a further derivative and going to a stationary point, 
one immediately deduces that $(m_0^2)_{Ki} X_a^i + (m_0^2)_{K \bar \jmath} \bar X_a^{\bar \jmath} = 0$,
which is the statement that the would-be Goldstone boson $\sigma_a$ is massless.

At this point, one may wonder whether one could perhaps get some other relevant metastability 
conditions by looking at the complex partners $\rho_a$ of the would-be Goldstone modes, which are
a priory physical scalar fields. In the limit of unbroken supersymmetry, these modes have the same 
masses as the vector bosons. Upon 
supersymmetry breaking, they however split, and if the scale of supersymmetry breaking is much 
larger than that of gauge symmetry breaking, this splitting may become larger than the average 
mass of the multiplet and give rise to tachyons. A priori, there is no obstruction against making 
the gauge symmetry breaking scale much larger than the scale of supersymmetry breaking, thereby
avoiding that some of these states become tachyonic. However, in such a limit the effect of the 
gauging on the sGoldstino masses gets suppressed, and the potential benefits from the presence
of the vector multiplets disappear. A careful study may then perhaps unravel a limitation on how
much one may increase the sGoldstino masses through a gauging, coming from the danger that
these other states $\rho_a$ become tachyonic. However, we have not been able to find any 
simple result along this line of reasoning. We thus refrain from reporting here the rather 
complicated expression for the mass matrix of the fields $\rho_a$, which consists of the 
mass matrix of the vectors plus a series of terms that involve various tensors built out of
$X_a^i$ and its derivatives contracted with the auxiliary fields $F^i$ and $D^a$.

As a final remark on this issue, let us note that $F^i$ is orthogonal to $X_a^i$, as a 
consequence of the gauge invariance of the superpotential. This means that the sGoldstini 
$\varphi_\pm$ and the above complex partners of the would-be Goldstones $\rho_a$ 
actually probe the scalar mass matrix in two different sectors, the former orthogonal to $X_a^i$ and the 
latter parallel to $X_a^i$. Moreover, in the absence of supersymmetry breaking, these two 
sector are disentangled: the former describes the light chiral multiplets and the latter the 
heavy vector multiplets. However, it should also be noted that there is no guarantee that 
the would-be Goldstone modes $\sigma_a$ and their complex partners $\rho_a$ represent 
independent modes. Indeed, the number of linearly independent vectors in each of 
the two sets of  $\sigma_a^I = (X_a^i, \bar X_a^{\bar \imath})$ and 
$\rho_a^I = (i X_a^i, - i \bar X_a^{\bar \imath})$ equals the rank of the matrix 
of the scalar products within each set, which coincides with the symmetric gauge bosons 
mass matrix $2 g_{i \bar \jmath} X^i_{\t{(}a} \bar X_{b\t{)}}^{\bar \jmath} = (m_1^2)_{ab}$. 
On the other hand, the total number of linearly 
independent vectors in the full set containing both the $\sigma_a^I$ and the $\rho_a^I$ may 
be lower, because some of the $\sigma_a^I$ may be linear combinations of the $\rho_a^I$
and vice versa. It is given by the rank of a twice bigger matrix with diagonal blocks given by 
$2 g_{i \bar \jmath} X_{\t{(}a}^i \bar X_{b\t{)}}^{\bar \jmath} = (m_1^2)_{ab}$ and off-diagonal 
blocks given by 
$2 g_{i \bar \jmath} X_{\t{[}a}^i \bar X_{b\t{]}}^{\bar \jmath} = -i f_{ab}^{\;\;\,c} D_c$, which is 
also equal to twice the rank of the Hermitian matrix
$2 g_{i \bar \jmath} X_a^i \bar X_b^{\bar \jmath} = (m_1^2)_{ab} - i f_{ab}^{\;\;\,c} D_c$. 
Indeed, the existence of a complex null vector $\bar v^b$ for this matrix implies 
that $X_a^i v^a = \bar X_a^{\bar \imath} \bar v^a = 0$, and for each such null 
vector there are thus two linear relations between the $\rho_a$ and the $\sigma_a$ 
along the real directions ${\rm Re}\,v^a$ and ${\rm Im}\,v^a$: 
${\rm Re}\,v^a \rho_a = - {\rm Im}\,v^a \sigma_a$ and ${\rm Im}\,v^a \rho_a = {\rm Re}\,v^a \sigma_a$.
In such a situation, two combinations of the $\rho_a$ are then unphysical would-be Goldstone 
modes too.

The result derived above for the average sGoldstino mass represents the rigid limit of 
the one derived in \cite{GRS3}, generalized to arbitrary non-Abelian gauge groups.
In terms of the gravitino mass $m_{3/2}$ and the Planck mass $M_{\rm P}$, the cosmological
constant reads $V_{\rm S} = F^i \bar F_i + \frac 12 D^a D_a - 3\,m_{3/2}^2 M_{\rm P}^2$, and the 
averaged sGoldstino mass is
\bea
m^2_{\varphi} \b=\b R \,F^i \bar F_i + \big(S + M_{\rm P}^{-2}\big) D^a D_a 
+ \s{\frac 14} \, T\, \frac {(D^a D_a)^2\!\!}{F^i \bar F_i} 
+ \big(M^2 - 4\, m_{3/2}^2 \big) \frac {D^a D_a\!}{F^i \bar F_i} \nn \\
\b\;\b +\, 2\, m_{3/2}^2 \,.
\eea
We again see that the main feature of this result, namely the dependence 
on the curvatures $R$, $S$, $T$ and on the mass $M^2$, is also captured in the rigid 
limit, in which $m_{3/2} \to 0$ and $M_{\rm P} \to \infty$. As before, gravitational effects influence 
only quantitatively the result.

In this case the necessary condition for metastability does not become sufficient even 
if for a given K\"ahler potential $K$ one allows the superpotential $W$ to be adjusted. 
Indeed, the restriction of gauge invariance of $W$ implies that $W_i X_a^i = 0$, $W_{ij} X_a^j = - \partial_i X_a^j W_j$ 
and $W_{ijk} X_a^k =  - 2 \partial_{\t{(}i} X_a^k W_{j\t{)}k} - \partial_i \partial_j X_a^k W_k$.
This shows that at the stationary point $W_i$, $W_{ij}$ and $W_{ijk}$ cannot be freely tuned along 
the complex directions associated to the Killing vectors $X_a^i$. The real modes corresponding to 
these directions are the would-be Goldstone modes $\sigma_a$ and their complex partners $\rho_a$. 
The masses of the latter can therefore not be adjusted through their $F$-term part depending on $W$ 
and represent a left-over danger, whenever they are physical. These masses have however also a 
$D$-term part depending on the Killing potentials $K_a$, and tend to the vector bosons masses 
in the supersymmetric limit. This suggests that if one could somehow also allow the Killing potential 
$K_a$ to be adjusted, the metastability condition would become once again effectively sufficient. 
The extent to which one can imagine to do that is however clearly restricted, since $K_a$, on the 
contrary of $W$, does have some relation to the geometry defined by $K$. 

\section{N=2 models with hyper multiplets}
\setcounter{equation}{0}

Let us now consider the simplest case of $N=2$ theories with $n_{\cal H}$ hyper multiplets ${\cal H}^{k}$.
This is a particular case of $N=1$ theory with $n_{\rm C} = 2 n_{\cal H}$ chiral multiplets $Q^u$, with the 
particularity that it admits a second supersymmetry. The most general two-derivative Lagrangian is specified 
by a real K\"ahler potential $K$ and a holomorphic superpotential $W$, and in $N=1$ superspace it 
takes the usual form
\be
\mathcal{L} = \int \! d^4 \theta\, K(Q,\bar Q) + \int \! d^2 \theta\, W(Q) + {\rm h.c.} \,.
\label{actionhyp}
\ee
The existence of a second supersymmetry mixing different $N=1$ superfields implies strong additional 
restrictions on $K$ and $W$. To derive these restrictions, we shall follow \cite{HKLR} and construct 
systematically the most general form of the second supersymmetry.

The general form of the second non-manifest supersymmetry transformation can be parametrized 
as follows with a general complex function $\bar N^u$, a holomorphic function $X^u$ and a phase 
$s$ \cite{HKLR,BX,Kuzenko}:
\be
\hat \delta Q^u = \s{\frac 12} \bar D^2 \Big(\bar N^u(Q,\bar Q) (\hat \epsilon \theta + \hat {\bar \epsilon} \bar \theta) \Big)
- 2 i (s + \bar s) X^u(Q) \hat \epsilon \theta \,. \label{deltahath}
\ee
In order for this to correctly satisfy an $N=1$ supersymmetry subalgebra, more precisely
$[\hat \delta_1,\hat \delta_2] Q^u = -2i (\hat \epsilon_1 \sigma^\mu \hat{\bar \epsilon}_2
-\hat \epsilon_2 \sigma^\mu \hat{\bar \epsilon}_1) \partial_\mu Q^u$, one needs to impose 
some restrictions on the functions $\bar N^u$ and $X^u$. A straightforward computation 
shows that the required conditions are the following:
\bea
\a\a \partial_{\bar w} \bar N^u \partial_v N^{\bar w} = - \delta^u_v \,,\;\;
\partial_{\bar s} \partial_v \bar N^u \partial_{\bar t} \bar N^v 
- \partial_{\bar t} \partial_v \bar N^u \partial_{\bar s} \bar N^v = 0 \label{algN} \\
\a\a \partial_w X^u \partial_{\bar v} \bar N^w 
- \partial_{\bar v} (\partial_{\bar w} \bar N^u \bar X^{\bar w}) 
-  \partial_{\bar v} \partial_w \bar N^u X^w = 0 \,,  \label{algX}
\eea

Let us now check under what circumstances the Lagrangian (\ref{actionhyp}) is left invariant 
by a second supersymmetry of this general allowed form. One finds that this is the case provided that
\bea
\a\a \nabla_u N_v + \nabla_v N_u = 0 \,,\;\;  
\nabla_w (\nabla_u N_v) = 0 \,,\;\; 
\nabla_{\bar w} (\nabla_u N_v) = 0 \label{invN} \,, \\
\a\a X^u = i \bar s \nabla^u \bar N^v W_v \,, \;\; K_u X^u + K_{\bar u} X^{\bar u} = f + \bar f \label{invX} \,.
\eea
In these equations, we have used the K\"ahler metric to raise and lower indices,
and $f$ denotes an arbitrary holomorphic function of the chiral multiplets.

In order to clarify the geometrical meaning of the above restrictions, let us introduce
the following notation:
\be
\Omega_{uv} = \nabla_u N_v \,.
\ee
In terms of this quantity, the constraints (\ref{invN}) for the invariance of the action
imply that $\Omega_{uv}$ should be antisymmetric, covariantly constant and holomorphic.
Moreover, the first constraint (\ref{algN}) from the closure of the algebra implies a further 
constraint on the contraction of $\Omega_{uv}$ with its conjugate, while the second of 
(\ref{algN}) is automatically satisfied as a consequence of the holomorphicity of $\Omega_{u v}$.
One thus finds:
\bea
\a\a \Omega_{uv} = - \Omega_{vu} \,,\;\;
\nabla_w \Omega_{uv} = 0 \,,\;\;
\nabla_{\bar w} \Omega_{uv} = 0 \,, \label{Omega1} \\
\a\a \bar \Omega^u_{\;\;\bar w} \Omega^{\bar w}_{\;\;v} = - \delta^u_v \label{Omega2}\,.
\eea
It then follows that the K\"ahler manifold admits three complex structures, constructed 
out of $\Omega_{uv}$ as
\be
(J^1)^U_{\;\;V} = \left(\begin{matrix} 
0 \!\!&\!\! \bar \Omega^u_{\;\; \bar v} \smallskip\ \cr 
\Omega^{\bar u}_{\;\; v} \!\!&\!\! 0 \cr 
\end{matrix}\right) \,,\;\;
(J^2)^U_{\;\;V} = \left(\begin{matrix} 
0 \!\!&\!\! \,i \bar \Omega^u_{\;\; \bar v}\, \smallskip\ \cr 
\mbox{-}i \Omega^{\bar u}_{\;\; v} \!\!&\!\! 0 \cr 
\end{matrix}\right) \,,\;\;
(J^3)^U_{\;\;V} = \left(\begin{matrix} 
\,i \delta^u_v\, \!\!&\!\! 0 \smallskip\ \cr 
0 \!\!&\!\! \mbox{-}i \delta^{\bar u}_{\bar v} \cr 
\end{matrix}\right) \,,
\ee
which are covariantly constant and satisfy the quaternions algebra:
\bea
\a\a \nabla_U (J^x)^V_{\;\; W} = 0 \,, \\
\a\a (J^x)^U_{\;\;W} (J^y)^W_{\;\;V} = - \delta^U_V \delta^{xy} + \epsilon^{xyz} (J^z)^U_{\;\;V} \,.
\eea
This means that the K\"ahler manifold must actually be Hyper-K\"ahler \cite{AGF,BW}. 

Notice that the transformation functions $\bar N^u$ are implicitly determined by the 
quantity $\Omega_{uv}$ specifying the quaternionic structure. Indeed, compatibly with
all the properties listed above, one can write:
\be
\bar N^u = - \bar \Omega^{uv} (K_v + f_v) \,,\;\; K_u \bar N^u = K_u g^u \,.
\ee
The arbitrary holomorphic functions $f_v$ and $g^u = - \bar \Omega^{uv} f_v$ 
reflect the ambiguities related to K\"ahler transformations of $K$ and in the 
definition of $\bar N^u$.

Concerning the superpotential, we see that the basic object controlling its structure
is the holomorphic vector defined by (\ref{invX}):
\be
X^u = i \bar s \bar \Omega^{uv} W_v \,
\label{XW}.
\ee
The constraints (\ref{invX}) from the invariance of the action imply,
upon taking some derivatives, that $X^u$ is holomorphic and satisfies the Killing equation,
whereas the condition (\ref{algX}) coming from the closure of the algebra implies that it 
also satisfies a further Killing-like equation involving $\bar \Omega^u_{\;\; \bar v}$:
\bea
\a\a \nabla_{\bar w} X^u = 0 \,,\;\; \nabla_u X_{\bar v} + \nabla_{\bar v} \bar X_u = 0 \,, \label{KA} \\
\a\a \bar \Omega^u_{\;\; \bar w} \nabla_{\bar v} \bar X^{\bar w} - \bar \Omega^w_{\;\; \bar v} \nabla_w X^u = 0\,.\label{KB}
\eea
This shows that $X^u$ must actually be a triholomorphic Killing vector of the Hyper-K\"ahler manifold,
meaning that the Lie derivative along it of any of the three complex structures $J^x$ must vanish:
\be
({\cal L}_{X + \bar X} J^x)^U_{\;\; V} = 0 \,.
\ee
For $x=3$, this is simply the statement in the first of the relations (\ref{KA}) that it is holomorphic with 
respect to the complex structure that is already manifest from the beginning, whereas for $x=1,2$ it 
amounts to the additional relation (\ref{KB}), which guarantees that it is also holomorphic with respect
to the two additional complex structures.

Since $X^U$ is a triholomorphic Killing vector, it admits three different real Killing potentials $P^x$, 
one for each complex structure $J^x$ (no sum over $x$):
\be
X^U = (J^x)^U_{\;\;V} \nabla^V \!P^x \,.
%\nabla^U P^x = - (J^x)^U_{\;\;V} X^V \,.
\ee
Notice that the Killing potentials $P^x$ are only defined modulo constants, which are here irrelevant.
In complex coordinates one then finds $X^u = \Omega^u_{\;\; \bar v} \nabla^{\bar v} P^1 
=  i \Omega^u_{\;\; \bar v} \nabla^{\bar v} P^2 = i \nabla^u P^3$. We see that $-P^3$ corresponds 
to the standard real Killing potential for $X^u$ seen as holomorphic with respect to $J^3$. 
In addition, one may however also use $P^1$ and $P^2$ to form a complex Killing potential 
\be
P = - \s{\frac i2} (P^1\! + i P^2) \,, 
\ee
which has the property of being holomorphic with respect to $J^3$:
\be
\nabla_{\bar u} P = 0 \,.
\ee
We may then write $X^u = i \nabla^u P^3$ but also $X^u = i \bar \Omega^{uv} P_v$. Comparing with 
(\ref{XW}), we see that the superpotential can be identified with this holomorphic Killing potential
\cite{AGF2}, times the phase $s$:
\be
W = s P \,.
\ee
Note that the phase $s$ cannot be trivially eliminated by rescaling $X^u$ and $P$, because only a 
real rescaling of these quantities preserves their defining properties. 

Having constructed the most general model that is invariant under both the usual and the extra supersymmetries,
we may now compute the commutator of such transformations and check that it closes only on-shell and with a
non-trivial central charge related to the Killing vector $X^u$. Indeed, the superfield equations of motion read 
$\bar D^2 K_u - 4 W_u = 0$ and thanks to the properties of the tensor $\Omega^{uv}$ they imply that
$D^2 \bar N^u + 4 i s X^u = 0$. Using this equation, one then easily verifies that
$[\delta_1, \hat \delta_2] Q^u = - 2i (\bar s \epsilon_1 \hat \epsilon_2 - s \bar \epsilon_1 \hat {\bar \epsilon}_2) X^u$,
whose right-hand side is of the form
\be
\delta_{\rm c} Q^u = \alpha X^u(\Phi) \,.
\ee
This central charge transformation corresponds to a global symmetry of the theory. Indeed, 
$\delta_{\rm c} K = X^u K_u + \bar X^{\bar u} K_{\bar u} = f + f$ and 
$\delta_{\rm c} W = X^u W_u = i s X^u \Omega_{uv} X^v = 0$, as a consequence of 
the second of (\ref{invX}) and the first of (\ref{Omega1}). It follows that the Lagrangian 
(\ref{actionhyp}) is invariant.

It is worth emphasizing that it is possible to consider alternative versions of the second supersymmetry 
transformations, which look different but yield the same on-shell transformations. For instance, as 
explained in \cite{BX} one may add to the transformation (\ref{deltahath}) the trivial transformation
$\delta_{\rm t} Q^u = \frac 12 \bar \Omega^{u v} (\bar D^2 K_v - 4 s P_v) \hat \epsilon \theta$, which 
is a symmetry of the on-shell theory since $\bar \Omega^{uv}$ is antisymmetric and 
the parenthesis is proportional to the equations of motion of $Q^u$. One then obtains
$\hat \delta Q^u = \frac 12 \bar D^2 (\bar N^u \hat {\bar \epsilon} \bar \theta)
- 2 i \bar s X^u \hat \epsilon \theta$.

Before going on, let us summarize some important features of Hyper-K\"ahler manifolds that will 
be relevant in the following. First, notice that the properties (\ref{Omega1}) 
imply that $\bar \Omega^u_{\;\;\bar w} \partial_{\t{[}s} \Omega^{\bar w}_{\;\; t\t{]}} = 0$ and that the 
Christoffel symbols are entirely determined in terms of $\Omega^u_{\;\; \bar v}$ and its conjugate:
\be
\Gamma^u_{st} = - \bar \Omega^u_{\;\;\bar w} \partial_{\t{(}s} \Omega^{\bar w}_{\;\; t\t{)}} \,.
\ee
From this expression one may compute the Ricci tensor and show that it identically vanishes,
due to the above properties of $\bar \Omega^u_{\;\; \bar v}$:
\be
R_{u \bar v} = 0 \,.
\ee
Finally, the integrability condition associated to the differential constraint (\ref{Omega1}) 
implies that the Riemann tensor, which is also completely determined by $\Omega^u_{\;\; \bar v}$ 
and its conjugate, satisfies the following algebraic constraint:
\be
\Omega^{\bar w}_{\;\; \t{[}u} R_{v\t{]} \bar w s \bar t}  = 0 \,.
\label{idRiem1}
\ee
Using (\ref{Omega2}), this also implies
\be
R_{u \bar v s \bar t}  = - \Omega^{\bar n}_{\;\; u} \bar \Omega^m_{\;\; \bar v} R_{s \bar n m \bar t} 
= \Omega^{\bar n}_{\;\; u} \bar \Omega^m_{\;\; \bar v} \Omega^{\bar q}_{\;\; s} \bar \Omega^p_{\;\; \bar t} R_{p \bar n m \bar q} \,.
\label{idRiem2}
\ee

Let us also quote for later reference the following important property of the triholomorphic Killing 
vector $X^u$, which follows from eqs.~(\ref{KA}) and (\ref{KB}):
\be
\nabla_u X^u = 0 \,.
\label{tracelessh}
\ee

To sum up, we see that in order to get an $N=2$ model, the geometry must be Hyper-K\"ahler and the 
superpotential must be given by the holomorphic Killing potential defining a triholomorphic Killing vector 
associated to a central charge:
\be
{\cal L} = \int \! d^4 \theta \, K (Q,\bar Q) + \int \! d^2 \theta\, s P(Q) + {\rm h.c.} \,.
\ee
The component Lagrangian reads
\bea
\mathcal{L} \b=\b-g_{u \bar v} \, \partial_\mu q^u \partial^\mu \bar q^{\bar v}
-i g_{u \bar v} \, \chi^u \big(\partial\!\!\!/\bar \chi^{\bar v} + 
\Gamma^{\bar v}_{\bar s \bar t} \, \partial\!\!\!/ \bar q^{\bar s} \bar \chi^{\bar t} \big) 
- V_{\rm S} - V_{\rm F} \,,
\eea
where:
\bea
V_{\rm S} \b=\b g_{u \bar v} X^u \bar X^{\bar v}  \\
V_{\rm F} \b=\b \s{\frac i2} s\, \Omega_{uw} \nabla_{v} X^w \, \chi^u \chi^v + {\rm h.c.}
- \s{\frac 14} R_{u \bar v s \bar t} \, \chi^u \chi^s \bar \chi^{\bar v} \bar \chi^{\bar t} \,,
\eea

The first supersymmetry transformations are specified by the usual action of the supercharges 
on the superfields and act as follows on component fields:
\bea
\a\a \delta q^u = \s{\sqrt{2}}\, \epsilon \,\chi^u \,, \\
\a\a \delta \chi^u = \s{\sqrt{2}}\, \epsilon \,F^u + \s{\sqrt{2}}i \,\partial\!\!\!/ q^u \bar \epsilon \,. \label{dchi}
\eea
The value of the auxiliary fields $F^u$ is
\be
F^u = i \bar s\, \bar \Omega^u_{\;\;\bar v} \bar X^{\bar v} + \s{\frac 12} \Gamma^u_{st}\, \chi^s \chi^t \,.
\ee
The action of the second supersymmetry is obtained by computing the components of the superfield
expression (\ref{deltahath}). One finds:
\bea
\a\a \hat \delta q^u = - \s{\sqrt{2}}\, \bar \Omega^u_{\;\;\bar v} \, \hat {\bar \epsilon} \,\bar \chi^v \,, \\
\a\a \hat \delta \chi^u = \s{\sqrt{2}}\, \hat \epsilon \, \hat F^u
\!+\! \s{\sqrt{2}}\,\Gamma^u_{st} \bar \Omega^s_{\;\; \bar v}  \, \hat {\bar \epsilon} \, \bar \chi^{\bar v} \chi^t 
+ \s{\sqrt{2}}i\,\bar \Omega^u_{\;\;\bar v} \partial\!\!\!/ \bar q^{\bar v} \hat {\bar \epsilon} \label{dhatchi}\,. 
\eea
The quantity $\hat F^u$ is found to be given by 
\be
\hat F^u = \bar \Omega^u_{\;\; \bar v} \big(\bar F^{\bar v} + (s + \bar s) P^{\bar v}
\!-\! \s{\frac 12}\,\Gamma^{\bar v}_{\bar s \bar t} \,\bar \chi^{\bar s} \bar \chi^{\bar t} \big) = - i \bar s X^u \,.
\ee

The extension to supergravity is described in \cite{DFF,ABCDFFM}. It turns out that 
there is no obstruction in coupling a model of the above type 
to gravity. The main new feature is that there appears a non-trivial $SU(2)$ bundle 
over the scalar manifold with curvature proportional to $M_{\rm P}^{-2}$, and the manifold 
becomes Quaternionic-K\"ahler. In this setting, the fact that the scalar potential depends on a 
Killing vector can be understood as coming from a gauging, of the type described in 
\cite{Zachos} and involving the graviphoton $A_\mu^0$. To see how it works, it is convenient to 
rewrite $X^u$ in terms of some new $X_0^u$ with dimension $1$ rather than $2$,
by introducing some mass scale $\mu$ and defining $X^u = \s{\sqrt{2}} \mu X_0$. One may further 
promote the mass scale $\mu$ to a complex mass parameter including the arbitrary phase $s$
appearing in the supersymmetry transformations laws, $L^0 = - i s \mu$, and write:
\be
X^u = \s{\sqrt{2}} i \bar s\, X_0^u L^0 \,.
\ee
Correspondingly, one may rewrite the Killing potentials as $P^x = \s{\sqrt{2}}i \bar s\, P^x_0 L^0$, in such a 
way that $P = \s{\sqrt{2}}i \bar s\,P_0 L^0$. For simplicity, we set from now on $s=i$, corresponding to $L^0$ real,
but it is clear that an arbitrary $s$ and a complex $L^0$ can be easily restored. We then see that the 
Lagrangian obtained above coincides with 
the one that emerges by taking a suitable rigid limit of $N=2$ supergravity coupled to hyper multiplets 
with a gauging of the central charge by the graviphoton $A_\mu^0$, whose action involves 
$X_0^u$. The non-trivial superpotential of the rigid theory, which is the generalization of the 
mass terms for the hyper multiplets that are allowed already in renormalizable theories, is 
obtained in the double scaling limit in which the Planck scale is sent to infinity and the graviphoton 
coupling to zero, but in such a way that their product gives rise to a finite mass scale. 

Notice finally that the scalar potential can be rewritten in a more familiar way by switching to 
general real coordinates: $2 g_{u \bar v} X_0^u \bar X_0^{\bar v} = g_{UV} X_0^U X_0^V$. 
One gets then the same result as in \cite{DFF,ABCDFFM}, namely:
\bea
V_{\rm S} \b=\b g_{UV} X_0^U \bar L^0 X_0^V L^0 \,.
\eea

\subsection{Supertrace}

At a generic point in the scalar field space and for vanishing fermions, the auxiliary fields simplify to
\be
F^u = \bar \Omega^u_{\;\;\bar v} \bar X^{\bar v} \,.
\ee
The corresponding hatted quantities similarly simplify to
\be
\hat F^u = \bar \Omega^u_{\;\; \bar v} \bar F^{\bar v} = - X^u \,.
\label{Fhathyp}
\ee 
The mass matrix of the scalar fields is given by
\bea
\a\a (m_0^2)_{u \bar v} = \nabla_u X^w \nabla_{\bar v} \bar X_w - R_{u \bar v s \bar t} \, F^s \bar F^{\bar t} \,, \\
\a\a (m_0^2)_{uv} = - R_{u \bar s v \bar t} \Omega^{\bar s}_{\;\; m} \Omega^{\bar t}_{\;\; n} F^m F^n 
+ \Gamma_{uv}^t V_{{\rm S}t}  \,.
\eea
The mass matrix of the fermions is instead
\bea
\a\a (m_{1/2})_{uv} = - \Omega_{\t{(}uw} \nabla_{v\t{)}} X^w \,.
\eea
Recalling that Hyper-K\"ahler manifolds are Ricci-flat, one easily computes
\bea
\a\a {\rm tr} [m_0^2] = 2\, \nabla_u X^v \nabla^u \bar X_v  \,,\\
\a\a {\rm tr} [m_{1/2}^2] = \nabla_u X^v \nabla^u \bar X_v \,.
\eea
It follows that the supertrace of the mass matrix vanishes \cite{HKLR}:
\bea
{\rm str}[m^2] \b\equiv\b {\rm tr} [m_0^2] - 2\, {\rm tr} [m_{1/2}^2] \nn \\
\b=\b 0 \,. \label{strh}
\eea
This result also follows directly from (\ref{strc}) and the fact that Hyper-K\"ahler
manifolds are Ricci-flat.

\subsection{Metastability}

The possible vacua of the theory correspond to points in the scalar manifold that 
satisfy the stationarity condition $V_{{\rm S}u} = 0$. This reads:
\be
\Omega_{uw} \nabla_{v} X^w F^v = 0 \,.
\ee

On the vacuum one has $\delta \psi^u = \s{\sqrt{2}} \epsilon F^u$ and $\hat \delta \psi^u = \s{\sqrt{2}} \hat \epsilon \hat F^u$,
and the first and second supersymmetries are spontaneously broken respectively if some of the auxiliary fields 
$F^u$ or some of the $\hat F^u$ are non-vanishing. The order parameters are the norms of the two vectors formed 
out of these two types of quantities. Since $\hat F^u \hat {\bar F}_u = F^u \bar F_u$, these two norms
actually coincide and both define the scalar potential energy, in two equivalent ways emphasizing the two supersymmetries:
$V_{\rm S} = F^u \bar F_u = \hat F^u \hat {\bar F}_u$. In such a situation, there are then two massless Goldstini 
given by:
\be
\eta = \s{\sqrt{2}}\, \bar F_u \chi^u \,,\;\;
\hat \eta = \s{\sqrt{2}}\, \hat {\bar F}_u \chi^u \,. 
\ee
Indeed, the stationarity condition implies that these are both flat directions of the fermion mass matrix:
\be
m_\eta = 0 \,,\;\; m_{\hat \eta} = 0 \,.
\ee
In this case the two supersymmetries can only be broken simultaneously. This is due to the fact that the
conditions that $F^u$ and $\hat F^u$ vanish are equivalent, since they are related by the invertible relation 
(\ref{Fhathyp}). From the structure of the supersymmetry transformations, we see that the 
four would-be supersymmetric scalar partners of these fermionic modes, the sGoldstini, can be parametrized 
by the four independent real linear combinations that one can form with the two complex 
Goldstino vectors $\eta^u = \s{\sqrt{2}} F^u$ and $\hat \eta^u = \s{\sqrt{2}}\, \hat F^u$:
\bea
\varphi_+ \b=\b \bar F_u q^u + F_{\bar u} \bar q^{\bar u} \,,\;\;
\varphi_- = i \bar F_u q^u - i F_{\bar u} \bar q^{\bar u} \,, \\
\hat \varphi_+ \b=\b \hat {\bar F}_u  q^u + \hat F_{\bar u} \bar q^{\bar u} \,,\;\;
\hat \varphi_- = i \hat {\bar F}_u q^u - i \hat F_{\bar u} \bar q^{\bar u} \,.
\eea
The masses of these four scalar modes can now be computed by evaluating 
the scalar mass matrix along the directions $\varphi_+^U = (F^u, \bar F^{\bar u})$,
$\varphi_-^U = (i F^u, - i \bar F^{\bar u})$, $\hat \varphi_+^U = (\hat F^u, \hat {\bar F}^{\bar u})$ 
and $\hat \varphi_-^U = (i \hat F^u, -i\hat {\bar F}^{\bar u})$, and dividing by the length of 
these vectors, which is $2 F^u \bar F_u = 2 \hat F^u \hat {\bar F}_u$. 
Notice that $F^u$ and $\hat F^u$ are orthogonal, $\hat F^u \bar F_u = 0$, and should thus 
lead to two independent informations. 

Viewing the theory as an $N=1$ theory with $F$ breaking, the first pair of masses is given by 
eq.~(\ref{m+-F}), with $R$ given by (\ref{RF}). The constraints imposed by 
the fact that the geometry is Hyper-K\"ahler do not substantially simplify neither the stationarity 
condition nor the form of the curvature at a stationary point, and one still has:
\be
R = \mbox{generically non-zero} \,.
\ee

Coming back to the $N=2$ picture, one may compute more explicitly all the four masses. 
After using the stationarity condition to simplify the results, one obtains:
\bea
m_{\varphi_\pm}^2 \b=\b (R \pm R_\Delta) \,F^u \bar F_u \,, \\
m_{\hat \varphi_\pm}^2 \b=\b (\hat R \pm \hat R_\Delta) \,F^u \bar F_u \,.
\eea
The terms involving the quantity $R$ and $\hat R$ come from the contributions of the Hermitian block 
$(m_0^2)_{i \bar \jmath}$ of the mass matrix, whereas the terms involving $R_\Delta$ and $\hat R_{\Delta}$
correspond to the contributions from the complex block $(m_0^2)_{i j}$. In this case,
these quantities are all related to sectional curvatures, and one finds
\bea
\a\a R = - \frac {R_{u \bar v s \bar t} F^u \bar F^{\bar v} F^s \bar F^{\bar t}}{(F^w \bar F_w)^2} \,, \\
\a\a R_\Delta = - \frac {R_{u \bar v s \bar t} F^u \hat {\bar F}^{\bar v} F^s \hat {\bar F}^{\bar t}}{2 (F^w \bar F_w)^2} + {\rm h.c.} \,, \\
\a\a \hat R = - \frac {R_{u \bar v s \bar t} \, \hat F^u \hat {\bar F}^{\bar v} F^s \bar F^{\bar t}}{(F^w \bar F_w)^2} \,, \\
\a\a \hat R_\Delta = - \frac {R_{u \bar v s \bar t} \, \hat F^u \hat {\bar F}^{\bar v} \hat F^s \hat {\bar F}^{\bar t}}{(F^w \bar F_w)^2} \,.
\eea
It then follows that
\bea
\a\a m^2_{\varphi} \equiv \s{\frac 12} \big(m_{\varphi_+}^2 + m_{\varphi_-}^2\big) = R \, F^u \bar F_u \,, \label{meth1} \\
\a\a \hat m^2_{\hat \varphi} \equiv \s{\frac 12} \big(m_{\hat \varphi_+}^2 + m_{\hat \varphi_-}^2\big) = \hat R \, F^u \bar F_u \label{meth2} \,.
\eea
This represents exactly the same type of information as in the case of $N=1$ theories with chiral 
multiplets, but once for each supersymmetry. 

The crucial sharpening in the necessary conditions for metastability comes now when one takes 
into account that the scalar manifold is not only K\"ahler but actually Hyper-K\"ahler. From
(\ref{idRiem2}) it follows indeed that:
\be
\hat R = - \hat R_\Delta = - R \,.
\ee
The four sGoldstino masses then simplify to:
\bea
m_{\varphi_\pm}^2 \b=\b (R \pm R_\Delta) \,F^u \bar F_u \,, \\
m_{\hat \varphi_+}^2 \b=\b 0 \,,\;\; m_{\hat \varphi_-}^2 = - 2\, R \,F^u \bar F_u \,.
\eea
This finally leads to the following results:
\bea
\a\a m^2_{\varphi\,{\rm sing}} \equiv m^2_{\hat \varphi_+} = 0 \;, \\
\a\a m^2_{\varphi\,{\rm trip}} \equiv \s{\frac 13} \big(m^2_{\varphi_+} \!+ m^2_{\varphi_-} \!+ m^2_{\hat \varphi_-}\big) = 0 \,.
\label{meth}
\eea
The first of these implies that there is always a massless mode, which can be interpreted as the Goldstone
boson of the spontaneously broken central charge symmetry. The second implies instead that 
there generically occurs at least one tachyonic mode.

The above results can be made more transparent by switching to more general real coordinates 
and exploiting the $SU(2)$ symmetry rotating the three complex structures $(J^x)^U_{\;\; V}$. More
precisely, the four sGoldstini can be organized as a singlet $\varphi_0^U = X^U$
plus a triplet $\varphi_{x}^U = (J^{x})^U_{\;\; V} X^V$, so that modulo irrelevant factors 
$\varphi_0^U = \hat \varphi_+^U$
and $\varphi_1^U = \varphi_+^U$, $\varphi_2^U = \varphi_-^U$, $\varphi_3^U = \hat \varphi_-^U$.
One then has $m_{\varphi_0}^2 = 0$, corresponding again to the Goldstone mode of the spontaneously
broken central charge symmetry, and $\sum_x m^2_{\varphi_x} = 0$, corresponding to an $SU(2)$ 
invariant sum rule on the masses of the remaining triplet of sGoldstini. More precisely, one finds:
\bea
\a\a m_{\varphi_0}^2 = 0 \,, \\
\a\a m_{\varphi_x}^2 = - 2\,R_x\,F^u F_u \,,
\eea
where $R_x$ denotes the holomorphic sectional curvature defined by the complex structure 
$(J^x)^U_{\;\;V}$ and the direction $X^U$:
\be
R_x = \frac {R_{U V M N} X^U (J^x X)^V X^M (J^x X)^N}{(X^K \bar X_K)^2} \,.
\ee
Indeed, one easily verifies that $R_1 = - \frac 12 (R + R_\Delta)$, $R_2 = - \frac 12 (R - R_\Delta)$ and 
$R_3 = R$. Moreover, the result (\ref{meth}) is now seen to descend directly from the integrability condition 
of the covariant constancy of the three complex structures, which reads 
$\Omega^{\bar w}_{\;\; \t{[}u} R_{v\t{]} \bar w s \bar t}  = 0$ and implies the following sum rule:
\be
{\sum}_x R_x = 0 \,. \label{sumrule}
\ee

Summarizing, besides the $N=1$ information on two of the sGoldstini, which implies that 
$m_{\varphi_1}^2 \!+ m_{\varphi_2}^2 = 2\, R \, F^u \bar F_u$, there is a further information on the other two sGoldstini 
coming from the second supersymmetry and which implies that $m_{\varphi_0}^2 = 0$,
corresponding to the Goldstone mode associated to the spontaneous breaking 
of the central charge symmetry, and $m_{\varphi_3} = - 2\, R\, F^u \bar F_u$. It follows that one 
of the sGoldstini always has a non-positive square mass, independently of the sign of $R$.
It should be emphasized that the $N=1$ metastability condition is recovered through 
the average of the sGoldstino masses associated to the first and second non-canonical 
complex structures, and not through the sGoldstino mass associated to the third canonical 
complex structure, which has instead the opposite sign. 

The above results are the rigid limit of the results obtained in \cite{GLS} for the supergravity case. 
The cosmological constant reads $V_{\rm S} = F^u \bar F_u - 3\, m_{3/2}^2 M_{\rm P}^2$ and the 
relevant combination of sGoldstino masses is
\bea
\a\a m^2_{\varphi\,{\rm trip}} = - 2\,M_{\rm P}^{-2} \, F^u \bar F_u + \s{\frac {16}3}\,m_{3/2}^2 \label{methlocal} \,. 
\eea
We see again that the main features of this result are also captured in the rigid limit, 
in which $m_{3/2} \to 0$ and $M_{\rm P} \to \infty$. Gravitational effects influence 
only quantitatively the result. For the triplet sGoldstino, the first term partly arises from the fact 
that in the local case the scalar manifold is Quaternionic-K\"ahler, rather than Hyper-K\"ahler, 
and the sum rule (\ref{sumrule}) is deformed due to the $SU(2)$ curvature of order 
$M_{\rm P}^{-2}$ characterizing these manifolds. The singlet sGoldstino, on the other hand,
is unphysical in the local case, the corresponding degree of freedom being eaten by the 
graviphoton. But in the limit defined by the double scaling in which $M_{\rm P} \to \infty$ 
and $g \to 0$ with $g M_{\rm P} \to$ finite, this becomes the physical massless Goldstone 
boson of the spontaneously broken central charge global symmetry.
This clarifies the rigid limit interpretation of the result of \cite{GLS}. It also allows to 
check their structure and their normalization by comparing them with the corresponding 
result found here. By doing so, one verifies in particular that the sectional curvatures 
must appear with opposite signs in the $N=1$ and the $N=2$ sGoldstino masses. This is 
related to the sum rule $R_1 + R_2 = - R_3 + {\cal O} (M_{\rm P}^{-2})$ holding on the 
three holomorphic sectional curvatures. One however also sees that the $N=2$ result
of \cite{GLS} must be wrong by a factor of $2$ in its dependence on the curvature, whereas
the sign is correct. We believe it may simply miss an overall factor of $2$ in its normalization,
which we have included in (\ref{methlocal}).
 
In this case it is not clear to what extent the necessary condition for metastability could
be made sufficient by allowing a tuning. Indeed, from the $N=1$ perspective the 
superpotential $W$ is not arbitrary but rather related to an isometry of the geometry 
defined by $K$. This substantially restricts the freedom to adjust it.
 
\section{N=2 models with vector multiplets}
\setcounter{equation}{0}

Let us continue by considering the case of $N=2$ theories with $n_{\cal V}$ vector multiplets ${\cal V}^{a}$.
This is a particular case of $N=1$ theory with $n_{\rm C} = n_{\cal V}$ chiral multiplets $\Phi^i$ plus 
$n_{\rm V} = n_{\cal V}$ vector multiplets $V^a$. The most general two-derivative Lagrangian is 
specified by a real K\"ahler potential $K$, a holomorphic superpotential $W$, a holomorphic gauge 
kinetic function $f_{ab}$, some holomorphic Killing vectors $X_a^i$ and some real Fayet-Iliopoulos 
constants $\xi_a$, and in $N=1$ superspace it reads
\be
\mathcal{L} = \int \! d^4 \theta\, \Big[K(\Phi,\bar \Phi,V) + \xi_a V^a \Big]
+ \int \! d^2 \theta\, \Big[W(\Phi) + \s{\frac 14}\, f_{ab} (\Phi) \,W^{a \alpha} W_\alpha^b \Big] + {\rm h.c.} \,.
\label{Lvec}
\ee
The existence of a second supersymmetry mixing different $N=1$ superfields implies further strong 
restrictions on $K$, $W$, $f_{ab}$ and $X_a^i$. To work out these restrictions, we follow again the logic of 
\cite{HKLR}, with some additional ingredients taken from \cite{ADM} (see also \cite{AADT}) to obtain 
the most general allowed superpotential, and also some generalization to make the formulation
covariant under general field reparametrizations.

The general form of the second supersymmetry can be parametrized in terms of two holomorphic 
functions $f^i_a$ and $L^a$ plus some complex constants $m^a$, and takes the following 
form:\footnote{The transformation (\ref{deltahatv2}) implies that 
$\hat \delta W^a = \s{\frac i{\sqrt{8}}}\, \hat \epsilon\, \bar D^2 \bar L^a(\bar \Phi) 
+ \s{\sqrt{2}}\, \partial \!\!\!/ L^a(\Phi)\, \hat {\bar \epsilon} + {\cal O}(V) + 4\, m^a\, \hat \epsilon$}
\bea
\a\a \hat \delta \Phi^i = \s{\sqrt{2}} i f_a^i(\Phi)\, \hat \epsilon\, W^a \,, \label{deltahatv1} \\[1mm]
\a\a \hat \delta V^a = - \s{\sqrt{2}} i \big(\bar L^a (\bar \Phi) - i\, f_{bc}^{\;\;\,a} \bar L^b(\bar \Phi) V^c + {\cal O}(V^2)
+ \s{\sqrt{8}}i\, m^a \bar \theta^2\big) \hat \epsilon\, \theta + {\rm h.c.} \label{deltahatv2}\,.
\eea
In order for this to correctly satisfy an $N=1$ supersymmetry subalgebra, more precisely 
$[\hat \delta_1,\hat \delta_2] \Phi^i = -2i (\hat \epsilon_1 \sigma^\mu \hat{\bar \epsilon}_2
-\hat \epsilon_2 \sigma^\mu \hat{\bar \epsilon}_1) \partial_\mu \Phi^i$ and 
$[\hat \delta_1,\hat \delta_2] V^a = -2i (\hat \epsilon_1 \sigma^\mu \hat{\bar \epsilon}_2
-\hat \epsilon_2 \sigma^\mu \hat{\bar \epsilon}_1) \partial_\mu V^a$, one needs to impose 
some relation between the functions $f^i_a$ and $L^a$. A straightforward computation shows 
that one just needs to require that:
\bea
\a\a f^i_a \partial_i L^b = \delta_a^b \,,\;\; f^i_a \partial_j L^a = \delta^i_j \label{algf} \,.
\eea

The invariance of the action defined by (\ref{Lvec}) under this second supersymmetry is 
instead guaranteed by the following constraints, where $M_a$ and $f_a$ denote arbitrary 
holomorphic functions and $e_a$ some complex constants: 
\bea
\a\a f_{ab} = - i f_a^i \, \partial_i M_b = - i f_b^i \, \partial_i M_a \,,\;\; 
f^i_a K_i \big|_{V=0}  - \s{\frac 12} f_{ab} \bar L^b - \s{\frac i2} \bar M_a = f_a \,, \label{invM} \\
\a\a W_i f^i_a = \s{\sqrt{2}} \big(e_a + i f_{ab} m^b\big) \,,\;\; X_a^i = f^i_b f_{ac}^{\;\;\;b} L^c \,, \label{WXai} \\[1mm]
\a\a \xi_a, e_a, m^b = 0\;\mbox{whenever}\; f_{bc}^{\;\;\,a} \neq 0 \,,\;\; i f_{ad} f_{bc}^{\;\;\; d} L^c = - f_{ba}^{\;\;\; c} M_c \,. 
\label{gaugeinv}
\eea

To find out the geometrical meaning of the above constraints, we need first of all to interpret the meaning 
of the holomorphic functions $L^a$ appearing in the transformation laws and the holomorphic functions 
$M_a$ parametrizing the constraints put by the invariance of the action. Concerning $L^a$, it is natural 
to think of them as representing a general reparametrization of the original fields $\Phi^i$. One can then 
define the Jacobian matrix of this transformation:
\be
f_i^a = \nabla_i L^a \,.
\ee
The constraints (\ref{algf}) from the closure of the algebra then imply that this Jacobian matrix is invertible 
and that the functions $f_a^i$ are given by the inverse of this matrix:
\be
f^i_a = (f^{\mbox{-}1})^i_a \,.
\ee
Concerning $M_a$, we may similarly introduce the matrix
\be
h_{ai} = \nabla_i M_a \,,
\ee
and denote its inverse by
\be
h^{ai} = (h^{\mbox{-}1})^{ai} \,.
\ee
The two constraints (\ref{invM}) coming from the invariance of the action then imply the 
following relations for the gauge kinetic function $f_{ab}$ and the K\"ahler metric $g_{i \bar \jmath}$,
where $h_{ab}$ denotes the real part of $f_{ab}$:
\bea
\a\a f_{ab} = -i f_a^i h_{ib} = - i f_b^i h_{ia} \,, \\
\a\a g_{i \bar \jmath} = h_{ab} f^a_i \bar f^b_{\bar \jmath} \,.
\eea
We now observe that the first of the relations (\ref{invM}) can be rewritten in terms of $L^a$ and $M_a$ 
as $f_{ab} = -i \partial M_b/\partial L^a = -i \partial M_a/\partial L^b$, and implies thus 
that modulo some irrelevant constants the functions $M_a$ must be the gradients with 
respect to the functions $L^a$ of some holomorphic function $M$:
\be
M = \mbox{holomorphic prepotential} \,.
\ee
In other words, this means that the index $a$ in $M_a$ can be interpreted as 
the derivative with respect to $L^a$. It finally follows that the K\"ahler potential 
and the gauge kinetic function are both determined by the prepotential $M$ 
and read: 
\bea
\a\a K = \s{\frac i2} \,\big(\bar M_a L^a - \bar L^a M_a \big) + {\cal O}(V) 
= \s{\frac i2} \,\big((\bar M\, e^{-2 V})_a L^a - (\bar L\, e^{-2 V})^a M_a \big) \,, \label{Kpre}\\
\a\a f_{ab} = - i M_{ab}  \label{fpre} \,.
\eea
This is the statement that the geometry is Special-K\"ahler \cite{dWvP,Many,dWLvP,S}, 
with $L^a$ and $M_a$ playing the roles 
of the electric and magnetic components of the symplectic sections.

Concerning the superpotential, the constraints (\ref{WXai}) and (\ref{gaugeinv}) from the invariance of the action 
imply that it is restricted to be a linear combination of the electric and magnetic sections 
$L^a$ and $M_a$ corresponding to Abelian factors, with complex coefficients 
$e_a$ and $m^a$:
\be
W = \s{\sqrt{2}} \big(e_a L^a + m^a M_a \big) \,.
\ee
This superpotential for the $N=1$ chiral superfields $\Phi^i$, which is linear in the sections, 
represents the $N=2$ completion of the possibility of having a linear Fayet-Iliopoulos term 
for the $N=1$ vector superfields $V^a$. More precisely, the term linear in $L^a$ is trivially 
invariant on its own, thanks to the fact that the natural partners of the vector superfields $V^a$ 
under the second supersymmetry are the sections $L^a$, in the sense that 
$\hat \delta L^a = \s{\sqrt{2}} i \hat \epsilon W^a$. On the other hand, the term in $M_a$ is 
non-trivially invariant, and its variation $\hat \delta M_a = - \s{\sqrt{2}} \hat \epsilon f_{ab} W^b$ 
is canceled by the extra variation of the vector kinetic term induced by the explicit shift in 
$\hat \delta W^a$ proportional to the coefficients $m^a$.

We see that the well-known symplectic structure of $N=2$ theories with only vector multiplets emerges 
quite naturally from this framework. Moreover, one automatically finds a coordinate-covariant formulation,
along the lines of \cite{CDF,CDFV}. For vanishing non-Abelian gauge couplings and vanishing Fayet-Iliopoulos 
parameters, the theory is invariant under a duality symmetry acting as symplectic transformations on 
the sections $(L^a,M_a)$. 

At this point, one may check that the two supersymmetries commute, meaning that there is no central 
charge in this case: $[\delta_1, \hat \delta_2] \Phi^i = 0$, $[\delta_1, \hat \delta_2] V^a = 0$. This means
that the full supersymmetry algebra closes off-shell.

The form of the gauge transformations leaving the action invariant is fixed by the expression 
(\ref{WXai}) that the Killing vector must take:\footnote{One 
also has $\delta_{\rm g} W^a = f_{bc}^{\;\;\;a} \Lambda^b W^c$.}
\bea
\a\a \delta_{\rm g} \Phi^i= f^i_a f_{bc}^{\;\;\; a} \Lambda^b L^c \,, \\
\a\a \delta_{\rm g} V^a = - \s{\frac i2} \big(\Lambda^a - \bar \Lambda^a) 
+ \s{\frac 12} f_{bc}^{\;\;\;a} \big(\Lambda^b + \bar \Lambda^b) V^c + {\cal O}(V^2)  \,.
\eea
This means that the sections $L^a$ must transform in the adjoint representation of the gauge group:
$\delta L^a = f_{bc}^{\;\;\; a} \Lambda^b L^c$.
The properties (\ref{gaugeinv}) then guarantee that the Lagrangian is gauge invariant. 
Indeed, the invariance of the K\"ahler potential requires that $\delta M_a = - f_{ba}^{\;\;\; c} \Lambda^b M_c$. 
But since $\delta M_a = M_{ad} \delta L^d$, this implies the constraint 
$M_{ad} f_{bc}^{\;\;\; d} L^c = - f_{ba}^{\;\;\; c} M_c$, which coincides with the second of (\ref{gaugeinv}).
The invariance of the gauge kinetic term further requires that 
$\delta f_{ab} = 2i f_{c\t{(}a}^{\;\;\;\; d} M_{b\t{)}d} \Lambda^c$. But since $\delta f_{ab} = - i M_{abe} \delta L^e$, 
this implies that $M_{abe} f_{cd}^{\;\;\; e} L^d = - 2 f_{c\t{(}a}^{\;\;\; d} M_{b\t{)}d}$. It is however straightforward 
to check that this relation automatically follows from the former constraint, by taking a further derivative.  
Finally, the invariance of the superpotential implies that it should vanish in the non-Abelian directions, 
corresponding to the first condition in (\ref{gaugeinv}).

Before going on, let us summarize some important results concerning Special-K\"ahler geometry. 
The basic objects characterizing such a geometry are the sections $L^a$ and the following 
holomorphic symmetric tensor, which is related to the third derivative of the prepotential $M$ 
\cite{CDFSK1,CDFSK2}:
\be
C_{ijk} = \s{\frac 12} M_{abc} f_i^a f_j^b f_k^c \,.
\ee
Indeed, the Christoffel symbols and the Riemann tensor are found to be given by the following 
expressions:
\bea
\a\a \Gamma^i_{jk} = \partial_j f_k^a f_a^i - i C_{jkl} \bar f^{ia} f_a^l\,, \label{GammaSK} \\
\a\a R_{i \bar \jmath p \bar q} = - C_{i p k} \bar C^k_{\;\, \bar \jmath \bar q} \,.
\label{RiemannSK}
\eea
From (\ref{GammaSK}) one then deduces the following basic relation underlying 
Special-K\"ahler geometry, out of which the expression (\ref{RiemannSK}) for the Riemann 
tensor emerges as the integrability condition:
\bea
\a\a \nabla_i f_j^a = i C_{ijk} \bar f^{ka} \,. \label{basicSK} 
\eea
From this it also follows that:
\bea
\a\a \nabla_{\t{[}i} C_{j\t{]}kl} = 0 \,.
\eea
One also easily finds
\bea
\a\a h_{abi} = - i C_{ijk} f^j_a f^k_b \,, \\
\a\a \nabla_i h_{abj} - 2\, h_{aci} h^{cd} h_{bdj} = - i \nabla_i C_{jkl} f^k_a f^l_b \,.
\eea

In addition to the above restrictions posed by the geometry, there are also a number of relations 
descending from the fact that the sections describing the scalar fields transform in the adjoint 
representation and the Killing vectors $X_a^i$ are rigidly fixed and given by the second of eq.~(\ref{WXai}).
Since the K\"ahler potential is strictly invariant, the real Killing potentials associated to these Killing vectors 
are determined by $K_a = - 2 i X_a^i K_i = 2 i \bar X_a^{\bar \jmath} K_{\bar \jmath} $, in such a way that
$X_a^i = \frac i2 \, g^{i \bar \jmath} \nabla_{\bar \jmath} K_a$. Using the second of (\ref{gaugeinv}) and its derivative, 
one then finds the following two equivalent expressions:
\bea
K_a \b=\b f_{ab}^{\;\;\; c} \big(L^b \bar M_c + \bar L^b M_c\big) \nn \\
\b=\b 2 i\, h_{ad} f_{bc}^{\;\;\; d} \bar L^b L^c \,.
\label{KSK}
\eea
From the expressions (\ref{WXai}) and (\ref{KSK}) one then derives the following identities:
\be
X_a^i L^a = 0 \,,\;\; 
X_a^i \bar L^a = - \s{\frac i2} \bar f^{ia} K_a \,,\;\; 
K_a L^a = 0 \,,\;\; K_a \bar L^a = 0 \,.
\label{contrSK}
\ee
The equivariance condition reads
\be
g_{i \bar \jmath} X_{\t{[}a}^i \bar X_{b\t{]}}^{\bar \jmath} = \s{\frac i4} f_{ab}^{\;\;\; c} K_c \,.
\ee
Moreover, as a consequence of the identity $M_{abe} f_{cd}^{\;\;\; e} L^d = - 2 f_{c\t{(}a}^{\;\;\; d} M_{b\t{)}d}$ 
implied by the transformation properties of the gauge kinetic function, one finds the following cyclic identity:
\be
X_a^i h_{bci} + X_b^i h_{cai} + X_c^i h_{abi} = 0 \,.
\label{cyclic}
\ee
Notice finally that using (\ref{basicSK}) one deduces that 
$\nabla_i X_{a \bar \jmath} = f_i^b \bar f_{\bar \jmath}^c \big( f_{ab}^{\;\;\; d} h_{dc} + X_a^k h_{bck} \big)$,
and using then (\ref{cyclic}) and the fact that $f_{ab}^{\;\;\;b} = 0$, one arrives at the following identity:
\be
\nabla_i X_a^i = - 2 X_b^k h^{bc} h_{cak} \,.
\label{tracelessv}
\ee

To summarize, the Lagrangian takes the following general form, after choosing the Wess-Zumino gauge:
\bea
\mathcal{L} \b=\b \int \! d^4 \theta\, \Big[K(\Phi,\bar \Phi) + \big(K_a(\Phi,\bar \Phi) + \xi_a\big) V^a 
+ 2\, g_{i \bar \jmath} (\Phi,\bar \Phi) X^i_a(\Phi) \bar X^{\bar \jmath}_b (\bar \Phi) V^a V^b \Big] \nn \\
\b\;\b + \int \! d^2 \theta\, \Big[\s{\sqrt{2}} \big(e_a L^a(\Phi) + m^a M_a(\Phi) \big)
- \s{\frac i4}\, M_{ab} (\Phi) \,W^{a \alpha} W_\alpha^b \Big] + {\rm h.c.} \,.
\eea
One may now verify more explicitly that this is invariant under the second supersymmetry, by 
retaining only terms at most linear in the vector multiplets in eqs.~(\ref{deltahatv1}) and (\ref{deltahatv2}).
In components, this gives
\bea
\mathcal{L} \b=\b - g_{i \bar \jmath} \, D_\mu \phi^i D^\mu \bar \phi^{\bar \jmath}
- \s{\frac 14} h_{ab} \, F^a_{\mu\nu} F^{b\mu\nu} + \s{\frac 14} k_{ab} \, F^a_{\mu\nu}\tilde{F}^{b\mu\nu} 
-ig_{i \bar \jmath} \, \psi^i \big(D\!\!\!\!/\,\bar \psi^{\bar \jmath} + 
\Gamma^{\bar \jmath}_{\bar m \bar n} \, D\!\!\!\!/\, \bar \phi^{\bar m} \bar \psi^{\bar n} \big) \nn \\
\b\;\b - \s{\frac i2} h_{ab} \, \lambda^{a} D\!\!\!\!/\, \bar \lambda^b + {\rm h.c.}  
- \s{\frac i{\!\sqrt{2}}} C_{ijk} f^j_a f^k_b \, \lambda^a \sigma^{\mu \nu} \psi^i F^{b}_{\mu\nu} + {\rm h.c.} 
- V_{\rm S} - V_{\rm F} \,, 
\eea
where:
\bea
V_{\rm S} \b=\b 2 h^{ab} (e_a + i f_{ac} m^c) (\bar e_b - i \bar f_{bd} \bar m^d)
+ \s{\frac 18} \, h^{ab}(K_a \!+ \xi_a)(K_b \!+ \xi_b) \,, \label{VSvec} \\[-0.5mm]
V_{\rm F} \b=\b \s{\frac 12} \Big[\s{\sqrt{2}} i\, C_{ijk} \bar f^{ka} \big((e_a - i \bar f_{ab} m^b) \psi^i \psi^j 
+ (\bar e_a - i \bar f_{ab} \bar m^b)  f^i_b f^j_c \lambda^b \lambda^c \big) \nn \\
\b\;\b \hspace{13pt} + \s{\sqrt{8}} \big(\bar X_{ai} + \s{\frac 14} C_{ijk} f^j_a \bar f^{kb} (K_b \!+ \xi_b) \big) 
\psi^i \lambda^a \Big] + {\rm h.c.} \nn \\
\b\;\b - \s{\frac 14} R_{i \bar \jmath k \bar l} \, \big(\psi^i \psi^k \bar \psi^{\bar \jmath} \bar \psi^{\bar l}
+ f^i_a f^k_b \bar f^{\bar \jmath}_c \bar f^{\bar l}_d \, \lambda^a \lambda^b \bar \lambda^c \bar \lambda^d 
+ 2 f^k_a \bar f^{\bar l}_b \, \psi^i \lambda^a \bar \psi^{\bar \jmath} \bar \lambda^b \big) \nn \\
\b\;\b + \s{\frac 14} \Big[\big(i \nabla_i C_{j k l} + 2\, C_{ikm} C_{jln} f^m_c \bar f^{nc}\big) f^k_a f^l_b\, \psi^i \psi^j \lambda^a \lambda^b \nn \\[0.5mm]
\b\;\b \hspace{20pt} +\, C_{ikm} C_{jln}  f^m_c \bar f^{nc} f^k_a f^l_b \, \psi^i \lambda^a \psi^j \lambda^b 
\Big]+ {\rm h.c.} \,.
\eea

The first supersymmetry transformation laws involve not only the usual action of 
the supercharge, but also a compensating gauge transformation with superfield parameter
$\Lambda^{a} = 2 i \theta \sigma^\mu \bar \epsilon A^a_\mu + 2\theta^2 \bar \epsilon \bar \lambda^a$
needed to preserve the Wess-Zumino gauge choice. 
The additional gauge transformation turns the ordinary derivative 
appearing in $\delta \psi^i$ into a gauge-covariant derivative, and one finds
\bea
\a\a \delta \phi^i = \s{\sqrt{2}}\, \epsilon\, \psi^i \,, \\
\a\a \delta \psi^i = \s{\sqrt{2}}\, \epsilon\, F^i + \s{\sqrt{2}}i\,D\!\!\!\!/\, \phi^i \, \bar \epsilon \,, \label{dpsi}\\
\a\a \delta A_\mu^a = i \epsilon\, \sigma_\mu \bar \lambda^a - i \lambda^a \sigma_\mu \, \bar \epsilon \,, \\
\a\a \delta \lambda^a = i \epsilon\, D^a + \sigma^{\mu \nu} \epsilon\, F_{\mu \nu}^a \,. \label{dlambda}
\eea
The auxiliary fields $F^i$ and $D^a$ are given by
\bea
\a\a F^i = - \s{\sqrt{2}}\, \bar f^{ia} (\bar e_a - i \bar f_{ab} \bar m^b) + \s{\frac 12} \Gamma^i_{mn} \, \psi^m \psi^n 
+ \s{\frac i2} \bar C^i_{\;\,\bar m \bar n} \bar f^{\bar m}_a \bar f^{\bar n}_b \, \bar \lambda^a \bar \lambda^b \,, \\
\a\a D^a = - \s{\frac 12} h^{ab} (K_b \!+ \xi_b) - \s{\frac 1{\sqrt{2}}}\, C_{i j k} \bar f^{ja} f^k_b \, \psi^i \lambda^b + {\rm h.c.} \,.
\eea
The second supersymmetry transformation laws similarly involve not only (\ref{deltahatv1}), (\ref{deltahatv2}), 
but also a gauge transformation with superfield parameter 
$\hat \Lambda^{a} = - \s{\sqrt{8}} \theta \hat \epsilon \bar L^a - 2 \theta^2 \! f_i^a \hat \epsilon \psi^i$,
needed to preserve the Wess-Zumino gauge. The extra gauge transformation 
shifts the $D^a$ auxiliary field appearing in $\hat \delta \psi^i$ by $h^{ab} K_b$, and one finds
\bea
\a\a \hat \delta \phi^i = \s{\sqrt{2}}\, \hat \epsilon\, f^i_a \lambda^a \,, \\
\a\a \hat \delta \psi^i = \s{\sqrt{2}}\, \hat \epsilon\, \hat F^i
+ \s{\sqrt{2}}\, \partial_j f_a^i \psi^j (\hat \epsilon \lambda^a)
+ \sigma^{\mu \nu} \hat \epsilon\, f^i_a F_{\mu \nu}^a \,, \label{dhatpsi}\\
\a\a \hat \delta A_\mu^a = - i\, \hat \epsilon\, \sigma_\mu \bar f^a_{\bar \imath} \bar \psi^{\bar \imath} 
+ i f^a_i \psi^i \sigma_\mu\, \hat{\bar \epsilon} \,, \\[0.5mm]
\a\a \hat \delta \lambda^a = i \hat \epsilon \, \hat D^a + \s{\sqrt{2}}i\,f^a_i D\!\!\!\!/\, \phi^i \, \hat {\bar \epsilon} 
\label{dhatlambda}\,.
\eea
The quantities $\hat F^i$ and $\hat D^a$ appearing in these expressions are found to be given by
\bea 
\a\a \hat F^i = \s{\frac i{\sqrt{2}}} f^i_a \big(D^a + h^{ab} K_b\big) 
= \s{\frac i{\sqrt{8}}} \bar f^{ia} (K_a - \xi_a) + \text{ferm.} \,, \\
\a\a \hat D^a = - \s{\sqrt{2}} i\, \big(\bar f^a_{\bar \imath} \bar F^{\bar \imath} + \s{\sqrt{8}}i\, m^a 
- \s{\frac 12} \, \partial_{\bar \imath} f_{\bar \jmath}^a \bar \psi^{\bar \imath} \bar \psi^{\bar \jmath} \big)
= 2i\, h^{ab} (e_b - i \bar f_{bc} m^c) + \text{ferm.} \,.
\eea

It is clear from the form of these expressions that the vectors $(\lambda^a, f^a_i \psi^i)$ are doublets of the 
$SU(2)_R$ automorphism group of the $N=2$ supersymmetry algebra. In particular, the second supersymmetry 
transformation can be obtained by supplementing the first supersymmetry transformation with the non-trivial 
element of the center $Z_2$ of $SU(2)_R$, acting as $(\lambda^a, f^a_i \psi^i) \to (- f^a_i \psi^i, \lambda^a)$. 
The above transformation laws, derived by using an $N=1$ superfield approach, agree with those derived in 
a component approach in \cite{FIS1,FIS2,FIS3,FI} by imposing the above $Z_2$ invariance, in the special case
where $f_i^a = \delta_i^a$.

The extension to supergravity was developed in \cite{dWvP,dWLvP,DFF,ABCDFFM}. It presents again some 
subtleties related to those terms in the action that were not genuinely but accidentally invariant. More precisely, 
it turns out that models of the above type can be consistently coupled to gravity only if the coefficients of the 
Fayet-Iliopoulos terms and the electric and magnetic 
linear superpotentials satisfy some restrictions. Again, this is due to the fact that the trivial invariance of such terms 
in the rigid limit is spoiled by gravitational effects. The main new feature is that there appears a 
non-trivial $U(1)$ bundle over the scalar manifold with curvature proportional to $M_{\rm P}^{-2}$, and the manifold 
becomes Special-K\"ahler-Hodge. To spell out more precisely the restrictions that need to be imposed on the $N=2$
Fayet-Iliopoulos terms, let us set the complex magnetic constants to $0$:\footnote{For the inclusion of magnetic 
gaugings, see \cite{MAG1,MAG2,MAG3}.}
\be
m^a = 0 \,.
\ee
Let us furthermore parametrize the real Fayet-Iliopoulos constants $\xi_a$ and the complex electric constants 
$e_a$ in terms of a triplet of real constants $P_a^x$:
\bea
P_a^1 = 2\, {\rm Re} (e_a) \,,\;\; P_a^2 = 2 \,{\rm Im} (e_a)  \,,\;\; P_a^3 = \s{\frac 12} \xi_a \,.
\label{Pax}
\eea
It is quite common to introduce also a similar notation for the non-Abelian part of the Killing potential, which 
is however not a constant but a real function of the scalar fields, and behaves as a singlet:
\be
P_a^0 = - \s{\frac 12} K_a \,.
\label{Pa}
\ee
The statement is then that in supergravity the triplet of constants $P_a^x$ must satisfy a non-trivial
equivariance condition, and are thus constrained. More precisely, there is a non-trivial effect coming from 
an $SU(2)$ curvature, which is of order $M_{\rm P}^{-2}$ and is thus a genuine supergravity effect.
For Abelian factors, however, this is the only term that arises, and one then obtains a constraint that is 
independent of $M_{\rm P}^{-2}$ and survives in the rigid limit. This constraint on $N=2$ theories is
the analogue of the constraint on $N=1$ theories that the Fayet-Iliopoulos term can arise only under
the very special circumstance that it is associated to a gauged $U(1)_R$ symmetry, and it reads
\be
\epsilon^{xyz} P^y_a P^z_b = 0 \,.
\label{align}
\ee
This means that when interpreted as trivectors, the $P^x_a$ for the various values of $a$ must all be 
parallel. The general solution to this equivariance condition is then parametrized in terms of a single 
trivector $P^x$, whose direction defines a definite $U(1)_R$ subgroup of $SU(2)_R$, and some real 
coefficients $p_a$:
\be
P_a^x = p_a P^x \,.
\label{solalign}
\ee
Notice that in terms of the original coefficients, this restriction implies that besides having the $\xi_a$ real, 
one needs also the $e_a$ to have all the same phase $z$. We shall here allow for non-zero $\xi_a$, 
contrarily to what we did in the $N=1$ case, since as soon as $P^x$ is not zero, 
we are in the peculiar situation where a $U(1)_R$ symmetry is gauged when gravity is switched on. 
From now on, we will then 
restrict to theories of this type, admitting a consistent coupling to gravity, whereas we 
shall discard to other more peculiar possibility of gauging the whole $SU(2)_R$. In this situation, 
the superpotential takes the form $W = \s{\sqrt{2}}\,z |e_a| L^a$ and as a result it satisfies 
the following relation, descending from (\ref{basicSK}):
\be
\nabla_i W_j = i z^2 C_{ijk} \bar W^k \,.
\label{Wspec}
\ee

Notice finally that one can reshuffle the scalar potential (\ref{VSvec}) as follows. 
For the $F$-term part, we get $2 h^{ab} e_a \bar e_b = \frac 12 h^{ab} (P_a^1 P_b^1 + P_a^2 P_b^2)$.
For the $D$-term part,  three types of terms arise. First, we see from (\ref{contrSK}) that 
$\frac 18 h^{ab} K_a K_b = \frac 12 g_{i \bar \jmath} X_a^i \bar L^a \bar X_b^j L^b = \frac 12 h^{ab} P_a^0 P_b^0$. Next, 
$\frac 18 h^{ab} \xi_a \xi_b = \frac 12 h^{ab} P_a^3 P_b^3$. 
Finally, from the second of (\ref{KSK}) and the fact that $\xi_a$ is non-vanishing only for Abelian factors, it follows that 
$\frac 14 h^{ab} K_a \xi_b = 0$. The scalar potential can then be rewritten in the following form, which reproduces 
that of \cite{DFF,ABCDFFM}:
\bea
V_{\rm S} \b=\b \s{\frac 12} g_{i \bar \jmath} X_a^i \bar L^a \bar X_b^{\bar \jmath} L^b + \s{\frac 12} h^{ab} P_a^x P_b^x \nn \\
\b=\b \s{\frac 12} h^{ab} \big(P_a^0 P_b^0 + P_a^x P_b^x \big) \,.
\eea

\subsection{Supertrace}

At a generic point in the scalar field space and for vanishing fermions and vector fields, the auxiliary fields 
simplify to
\bea
\a\a F^i = - \s{\sqrt{2}}\,\bar f^{ia} \bar e_a  \,,\\
\a\a D^a = - \s{\frac 12} h^{ab} (K_b \! + \xi_b) \,.
\eea
The corresponding hatted quantities similarly simplify to 
\bea
\a\a \hat F^i = \s{\frac i{\sqrt{2}}} f^i_a \big(D^a + h^{ab} K_b\big) 
= \s{\frac i{\sqrt{8}}} \bar f^{ia} (K_a \! - \xi_a) 
= \hat F^i_{\t{\perp}} + \hat F^i_{\t{\parallel}} \,, \\
\a\a \hat D^a = - \s{\sqrt{2}} i\, \bar f^a_{\bar \imath} \bar F^{\bar \imath} 
= 2 i\, h^{ab} e_b \,.
\eea
The mass matrix of the scalar fields is given by
\bea
\a\a (m_0^2)_{i \bar \jmath} = - R_{i \bar \jmath k \bar l} \, 
\big(2\,F^k \bar F^{\bar l} + f^k_a \bar f^{\bar l}_b\, D^a D^b \big) + h^{ab} \bar X_{a i} X_{b \bar \jmath} \nn \\[0mm]
\a\a \hspace{45pt} +\, \s{\frac i2} \big(\nabla_i X_{a \bar \jmath} - 2 h^{bc} h_{abi} X_{c \bar \jmath} \big)\,D^a + {\rm h.c.} \,, \\[0mm]
\a\a (m_0^2)_{ij} = \s{\frac i2} \nabla_i C_{jkl} \, \big(2\, z^2 F^k F^l + f^k_a f^l_b D^a D^b \big) - h^{ab} \bar X_{a i} \bar X_{b j} \nn \\[0.5mm]
\a\a \hspace{45pt} +\, 2i h^{bc} h_{ab\t{(}i} \bar X_{cj\t{)}} D^a + \Gamma_{ij}^k V_{{\rm S}k} \,,
\eea
The mass matrix of the fermions reads instead
\bea
\a\a (m_{1/2})_{ij} = - i z^2 C_{ijk} \, F^k \,, \\[1mm]
\a\a (m_{1/2})_{ab} =  - i C_{ijk} f^i_a f^j_b \, F^k \,, \\[0mm]
\a\a (m_{1/2})_{ia} = \s{\sqrt{2}}\, \bar X_{ai} - \s{\frac 1{\!\sqrt{2}}} C_{ijk} f^j_a f^k_b\, D^b \,.
\eea
Finally, the mass matrix of the vectors is
\bea
\a\a (m_1^2)_{ab} = 2\, X_{\t{(}a}^i \bar X_{b\t{)}i} \,.
\eea
A straightforward computation gives
\bea
\a\a {\rm tr} [m_0^2] = 2\, R_{i \bar \jmath} \, 
\big(2\,F^i \bar F^{\bar \jmath} + f^i_a \bar f^{\bar \jmath}_b \, D^a D^b \big) + 2\, h^{ab} \bar X_{a i} X_b^i 
- 4 i\, h^{bc} h_{abi} X_c^i D^a + {\rm h.c.} \\[1mm]
\a\a {\rm tr} [m_{1/2}^2] = R_{i \bar \jmath} \, 
\big(2\,F^i \bar F^{\bar \jmath} + f^i_a \bar f^{\bar \jmath}_b \, D^a D^b \big)
+ 4\, h^{ab} \bar X_{ai} X_b^i - 2 i\, h^{ab} h_{bci} X_a^i D^c + {\rm h.c.} \,, \\[1mm]
\a\a {\rm tr} [m_1^2] = 2\, h^{ab} \bar X_{ai} X_b^i \,.
\eea
It follows that the supertrace of the mass matrix vanishes \cite{HKLR}:
\bea
{\rm str}[m^2] \b\equiv\b {\rm tr} [m_0^2] - 2\, {\rm tr} [m_{1/2}^2] + 3\, {\rm tr} [m_1^2] \nn \\[1mm]
\b=\b 0 \,.
\eea
This result also follows directly from (\ref{strcv}) and the properties that the Christoffel symbols are related to the 
derivative of the gauge kinetic function, the Ricci tensor to the contraction between two of these, and finally that 
the trace of the charge matrix satisfies the property (\ref{tracelessv}).

\subsection{Metastability}

The possible vacua of the theory correspond to points in the scalar manifold that 
satisfy the stationarity condition $V_{{\rm S}i} = 0$, which implies
\be
- \s{\frac i2} C_{ijk} \big(2\,z^2 F^j F^k + f^j_a f^k_b\,D^a D^b\big) + i \bar X_{a i} D^a = 0
\label{statvec}
\ee
The relation (\ref{relDFF}) between the values of the $F^i$ and $D^a$ auxiliary fields
can be simplified a bit by using the fact that $f_a^i \bar F_i$ vanishes for non-Abelian generators.
One finds
\be
i X_a^k h_{bck} \, f_i^b \bar f_{\bar \jmath}^c F^i \bar F^{\bar \jmath} - X_{\t{(}a}^i \bar X_{b\t{)} i} \, D^b 
+ \s{\frac 12} f_{ab}^{\;\;\;d} k_{dc} \,D^b D^c = 0 \,.
\label{DFFvec}
\ee

On the vacuum, one has $\delta \psi^i = \s{\sqrt{2}} \epsilon F^i$, $\delta \lambda^a = i \epsilon D^a$,
$\hat \delta \psi^i = \s{\sqrt{2}} \hat \epsilon \hat F^i$, $\hat \delta \lambda^a = i \hat \epsilon \hat D^a$,
and the first and second supersymmetries are spontaneously broken respectively if some of the auxiliary 
fields $F^i$, $D^a$ or some of the $\hat F^i$, $\hat D^a$ are non-vanishing.  The order parameters are given
by the norms of the two vectors built out of these two sets of quantities. 
Since $\hat F^i \hat {\bar F}_i = \frac 12 D^a D_a$ and $\frac 12 \hat D^a \hat D_a = F^i \bar F_i$, these 
two norms actually coincide and define again in two equivalent ways, emphasizing the two supersymmetries,
the scalar potential energy: 
$V_{\rm S} = F^i \bar F_i + \frac 12 D^a D_a = \hat F^i \hat {\bar F}_i + \frac 12 \hat D^a \hat D_a$.
In such a situation, there are then two massless Goldstini, associated to the two independent 
supersymmetries and given by:
\be
\eta = \s{\sqrt{2}} \bar F_i \psi^i + i D_a \lambda^a \,,\;\;
\hat \eta = \s{\sqrt{2}} \hat {\bar F}_i \psi^i + i \hat D_a \lambda^a \,,\;\;
\ee
In fact, one can verify that the stationarity condition and the gauge invariance of the superpotential 
imply that these are always flat directions of the fermion mass matrix:
\be
m_\eta = 0 \,,\;\; m_{\hat \eta} = 0 \,.
\label{resultmeta}
\ee
In the situation under consideration, the two supersymmetries can only be broken 
simultaneously.\footnote{The result (\ref{resultmeta}) actually holds true even in more general situations where the 
Fayet-Iliopoulos terms are not aligned and magnetic superpotentials are considered. In such a 
situation, partial supersymmetry breaking is known to be possible \cite{APT} (see also \cite{IZ, MOOP}). 
But in that case $F^i$ and $\hat F^i$ turn out to be parallel on the vacuum, and there is thus only 
one independent massless Goldstino.  In models compatible with gravity, 
on the other hand, partial supersymmetry breaking requires also the presence of hyper multiplets, whose 
presence can modify the alignment consistency condition \cite{FGP1,FGP2}.}
The sGoldstini are in this case linear combinations of scalars and vectors, but the 
relevant thing to look at is the projection along the scalar field space. One then gets four 
independent real linear combinations, corresponding to the projection of the complex Goldstino 
vectors $\eta^i = \s{\sqrt{2}} F^i$ and $\hat \eta^i = \s{\sqrt{2}} \hat F^i$:
\bea
\varphi_+ \b=\b \bar F_i \phi^i + F_{\bar \imath} \bar \phi^{\bar \imath} \,,\;\;
\varphi_- = i \bar F_i \phi^i - i F_{\bar \imath} \bar \phi^{\bar \imath} \,,\\
\hat \varphi_+ \b=\b \hat{\bar F}_i \phi^i + \hat F_{\bar \imath} \bar \phi^{\bar \imath} \,,\;\;
\hat \varphi_- = i \hat{\bar F}_i \phi^i - i \hat F_{\bar \imath} \bar \phi^{\bar \imath} \,.
\eea
The masses of these four scalar modes can now be computed by evaluating the scalar 
mass matrix along the directions $\varphi_+^I = (F^i, \bar F^{\bar \imath})$, 
$\varphi_-^I = (i F^i, - i \bar F^{\bar \imath})$, $\hat \varphi_+^I = (\hat F^i, \hat {\bar F}^{\bar \imath})$
and $\hat \varphi_-^I = (i \hat F^i, - i \hat {\bar F}^{\bar \imath})$, and dividing by the length of 
these vectors, which is $2 F^i \bar F_i$ for the first two and $2 \hat F^i \hat {\bar F}_i$ for the last 
two, with $F^i \bar F_i \neq \hat F^i \hat {\bar F}_i$. Notice however that $F^i$ and $\hat F^i$
are in general not orthogonal, and do thus not necessarily lead to two independent informations.
More precisely, one has $\hat F^i = \hat F^i_{\t{\perp}} + \hat F^i_{\t{\parallel}}$, where $\hat F_{\t{\perp}}^i$ is 
non-vanishing only in the non-Abelian case and orthogonal to $F^i$, whereas 
$\hat F_{\t{\parallel}}^i$ is non-vanishing whenever there are $N=1$ Fayet-Iliopoulos terms for some 
Abelian factors and is parallel to $F^i$ whenever the alignment condition on the $N=2$ Fayet-Iliopoulos 
terms is satisfied. 

Viewing the theory as an $N=1$ theory with $F$ and $D$ breaking, the first pair of masses is given by 
eq.~(\ref{m+-FD}), with $R$, $S$, $T$ and $M^2$ given by eqs~(\ref{RFD}), (\ref{SFD}), 
(\ref{TFD}) and (\ref{MFD}). But the constraints imposed by the fact that the geometry is Special-K\"ahler 
do in this case substantially simplify both the stationarity condition and the form of the curvatures, 
and there emerges a relation between the quantities $R$, $S$, $T$ and $M^2$ 
evaluated at a stationary point. This relations can be derived by solving for $C_{ijk} F^j F^k$ in the 
stationarity condition (\ref{statvec}) and taking its square norm. One then sees that the mixed terms 
drop out thanks to the properties implied by gauge invariance on the prepotential, and one deduces 
that 
\be
R\, F^i \bar F_i = \s{\frac 14}\, T\, \frac {(D^a D_a)^2}{F^i \bar F_i} + \s{\frac 12}\, M^2 \frac {D^a D_a}{F^i \bar F_i} \,.
\label{relRTM}
\ee

Coming back to the $N=2$ picture, one may compute more explicitly all the four masses and simplify
them by using the stationarity condition. To emphasize the important aspects of the results, we 
shall study separately the Abelian and non-Abelian cases.

\vskip 10pt
\noindent
{\bf Abelian case}
\vskip 5pt

\noindent
Consider first Abelian gauge groups. In this case
$X_a^i = 0$ and $K_a = 0$. One then has $\bar F_i = - \frac 1{\sq 2} f_i^a (P_a^1 + i P_a^2)$
and $\hat{\bar F}_i = \frac i{\sq 2} f_i^a P_a^3$, so that 
$F^i \bar F_i = \frac 12 h^{ab}(P^1_a P^1_b + P^2_a P^2_b)$ and
$\hat F^i \hat {\bar F}_i = \frac 12 h^{ab}P^3_a P^3_b$.

For simplicity, let us first study the situation where all the parallel Fayet-Iliopoulos parameters 
are rotated in the plane where $e_a \neq 0$ but $\xi_a = 0$. This implies that $F^i \neq 0$
but $\hat F^i = 0$. As a consequence, only the first pair of sGoldstino directions is well defined,
whereas the second pair is not. The first two sGoldstino masses are easily found to be given by:
\be
m_{\varphi_\pm}^2 = R \, F^i \bar F_i \pm \Delta \,.
\ee
In this expression, the quantity $R$ originates from the contribution from the Hermitian block
$(m^2_0)_{i \bar \jmath}$ of the mass matrix, whereas $\Delta$ encodes the contribution coming 
from the off-diagonal block $(m^2_0)_{ij}$. The former corresponds to a sectional curvature: 
\be
R = - \frac {R_{i \bar \jmath m \bar n}\, F^i \bar F^{\bar \jmath} F^m \bar F^{\bar n}}{(F^k \bar F_k)^2} \,.
\ee
It then follows that 
\bea
\a\a m_{\varphi}^2 \equiv \s{\frac 12} \big(m_{\varphi_+}^2 \!+ m_{\varphi_-}^2 \big) = R \, F^i \bar F_i \,. 
\eea
This result represents the informations associated to the first supersymmetry, to which a non-degenerate 
sGoldstino can be associated.

At this point, a sharp simplification does however occur when taking into account the form (\ref{RiemannSK}) 
implied for the Riemann tensor by the fact that the geometry is not only K\"ahler but actually Special-K\"ahler.
Indeed, we see that at a stationary point satisfying the stationarity condition $C_{ijk} F^j F^k = 0$, the 
sectional curvature $R$ actually vanishes. This corresponds to eq.~(\ref{relRTM}) applied to the present case:
\be
R = 0 \,.
\ee
The two sGoldstino masses then simplify to 
\be
m_{\varphi_\pm}^2 = \pm \Delta \,.
\ee
It finally follows that 
\bea
\a\a m_{\varphi}^2 = 0  \,. 
\eea

Let us now consider the more general situation where $e_a \neq 0$ and $\xi_a \neq 0$, where $F^i \neq 0$
and $\hat F^i \neq 0$. In this more general situation, both pairs of sGoldstini are well defined. However, we 
do not expect to get any additional information, since all the $\xi_a$ can be set to zero by an overall $SU(2)$
transformation, and we known that $V_{\rm S}$ is $SU(2)$ invariant. Nevertheless, it is instructive to see how
it works in this case. The four sGoldstino masses are found to be of the following form:
\bea
\a\a m_{\varphi_\pm}^2 = 2\, R\, F^i \bar F_i + 2\, R'\, \hat F^i \hat {\bar F}_i \pm \Delta \,, \\
\a\a m_{\hat \varphi_\pm}^2 = 2\, \hat R\, \hat F^i \hat {\bar F}_i + 2\, R'\, F^i \bar F_i \pm \hat \Delta \,.
\eea
In these expressions, the quantities $R$, $R'$ and $\hat R$ originate from the Hermitian block 
$(m^2_0)_{i \bar \jmath}$ of the mass matrix, whereas $\Delta$ and $\hat \Delta$ encode the 
contributions coming from the off-diagonal blocks $(m^2_0)_{ij}$. As usual, only the former have 
simple expressions, which are
\bea
\a\a R = - \frac {R_{i \bar \jmath m \bar n}\, F^i \bar F^{\bar \jmath} F^m \bar F^{\bar n}}{(F^k \bar F_k)^2} \,,\\
\a\a R' = - \frac {R_{i \bar \jmath m \bar n}\, F^i \bar F^{\bar \jmath} \hat F^m \hat {\bar F}^{\bar n}}
{(F^k \bar F_k)(\hat F^l \hat {\bar F}_l)} \,, \\
\a\a \hat R = - \frac {R_{i \bar \jmath m \bar n}\, \hat F^i \hat {\bar F}^{\bar \jmath} \hat F^m \hat {\bar F}^{\bar n}}
{(\hat F^k \hat {\bar F}_k)^2} \,.
\eea
Note that compared to the treatment of $N=1$ theories with $F$ and $D$ breaking of section 3,  
the quantities $R$, $R'$ and $\hat R$ introduced here correspond to the quantities $R$, $S$ and $T$,
whereas $F^i \bar F_i$ and $\hat F^i \hat {\bar F}_i$ correspond to $F^i \bar F_i$ and $\frac 12 D^a D_a$. 
Using the relation (\ref{relRTM}), we then see that the terms $R \, F^i \bar F_i$, $S \, D^a D_a$ and 
$T \, (D^a D_a)^2/(4\,F^i \bar F_i)$ in eq.~(\ref{m+-FD}) become respectively $R \,F^i \bar F_i$, $2 R' \hat F^i \hat {\bar F}_i$ 
and $R \,F^i \bar F_i$, and there is some simplification in the masses of the first pair of sGoldstini, 
whereas the mass of the new second pair of sGoldstini takes a similar expression with hatted and 
unhatted quantities exchanged. For the average of each pair of masses, one finds
\bea
\a\a m_{\varphi}^2 \equiv \s{\frac 12} \big(m_{\varphi_+}^2 \!+ m_{\varphi_-}^2 \big) 
= 2\, R\, F^i \bar F_i + 2 R'\, \hat F^i \hat {\bar F}_i  \,, \\
\a\a m_{\hat \varphi}^2 \equiv \s{\frac 12} \big(m_{\hat \varphi_+}^2 \!+ m_{\hat \varphi_-}^2 \big) 
= 2\, \hat R\, \hat F^i \hat {\bar F}_i + 2\, R'\, F^i \bar F_i  \,.
\eea
These results represent the informations associated to the two supersymmetries.
In the case of aligned Fayet-Iliopoulos terms, however, these two expressions 
should coincide and represent the same information, since $F^i$ and $\hat F^i$ 
are proportional to each other: $\hat F^i = i z (p_3 /\! \sqrt{p_1^2 + p_2^2}) F^i$.

The crucial simplification comes again from the form (\ref{RiemannSK}) of the 
Riemann tensor in Special-K\"ahler geometry. First, the stationarity condition
reads $C_{ijk} F^j F^k = \bar z^2 C_{ijk} \hat F^j \hat F^k$ and leads to a relation 
between $R$ and $\hat R$, which is just eq.~(\ref{relRTM}) applied to the present 
case. In addition, the alignment condition implies that 
$F^i \hat {\bar F}^{\bar \jmath} = - \bar z^2 \hat F^i {\bar F}^{\bar \jmath}$ and leads 
to a relation between $R'$ and $R$ or $\hat R$. The two relations are:
\be
R \, (F^i \bar F_i)^2 = \hat R \, (\hat F^i \hat {\bar F}_i)^2 = - R' (F^i \bar F_i) (\hat F^i \hat {\bar F}_i) \,.
\label{relRRhat}
\ee
The expressions for the four sGoldstino masses then simplify to
\bea
\a\a m_{\varphi_\pm}^2 = \pm \Delta \,, \\
\a\a m_{\hat \varphi_\pm}^2 = \pm \hat \Delta \,.
\eea
It finally follows that
\bea
\a\a m_{\varphi}^2 = 0 \,, \\
\a\a m_{\hat \varphi}^2 = 0 \,.
\eea
As expected, these two results coincide and it is clear that they represent the same information, 
since they are defined out of the two complex directions $F^i$ and $\hat F^i$, which are parallel.
There is thus really only one $SU(2)$-invariant information, stating that:
\be
m^2_{\varphi\,{\rm inv}} = 0 \,.
\ee

The above result represents the rigid limit of the result obtained in \cite{Many} for the 
supergravity case (see also \cite{FTV} for a derivation of the same result in the language 
of \cite{ABCDFFM}). The cosmological constant reads 
$V_{\rm S} = F^i \bar F_i + \frac 12 D^a D_a - 3\, m_{3/2}^2 M_{\rm P}^2$ 
and the average sGoldstino mass is
\be
m^2_{\varphi\,{\rm inv}} = - 2\, M_{\rm P}^{-2} \big(F^i \bar F_i + \s{\frac 12} D^a D_a\big) + 6 \, m_{3/2}^2 \,. 
\ee
Again, we see that the main feature of this result, namely the fact that it is independent of
the curvature, is also captured in the rigid limit, in which $m_{3/2} \to 0$ and $M_{\rm P} \to \infty$. 
Gravitational effects influence only quantitatively the result, making it negative instead of zero 
in the case of positive cosmological constant. 

\vskip 10pt
\noindent
{\bf Non-Abelian case}
\vskip 5pt

\noindent
Consider next non-Abelian gauge groups. In this case $X_a^i \neq 0$ and $K_a \neq 0$.
Then $\bar F_i  = - \frac 1{\sq 2} f_i^a (P_a^1 + i P_a^2)$ and 
$\hat{\bar F}_i = \frac i{\sq 2} f_i^a (P_a^3 + P_a^0)$, so that 
$F^i \bar F_i = \frac 12 h^{ab}(P^1_a P^1_b + P^2_a P^2_b)$
and $\hat F^i \hat {\bar F}_i = \frac 12 h^{ab}(P^3_a P^3_b + P_a^0 P_b^0)$.

As before, let us consider first the case where all the parallel Fayet-Iliopoulos terms are in the plane 
corresponding to $e_a \neq 0$ and $\xi_a = 0$. One then has $F^i \neq 0$ and $\hat F^i \neq 0$,
but whereas the first is truly generic the second is in fact related to the Killing vectors,
$\hat F^i = - \frac 1{\sq 2} X_a^i \bar L^a$, and this brings up some substantial simplifications.
In such a situation, all the four sGoldstini are well defined and their masses are found to be given by the following 
expressions, after using the stationarity conditions and all the relations descending from gauge invariance:
\bea
m_{\varphi_\pm}^2 \b=\b 2\, R\, F^i \bar F_i + 2\, R'\, \hat F^i \hat{\bar F}_i + M^2 \frac {\hat F^i \hat{\bar F}_i}{F^j \bar F_j} \pm \Delta \,, \\
m_{\hat \varphi_\pm}^2 \b=\b 0\,.
\eea
In these expressions, the quantities $R$, $R'$ and $M^2$ emerge from the contribution of the diagonal 
block $(m_0^2)_{i \bar \jmath}$ of the mass matrix, whereas $\Delta$ encodes the contribution from
the off-diagonal block $(m^2_0)_{ij}$. The quantities $R$, $R'$ and $M^2$, together with the quantity 
$\hat R$ introduced for later use, are given by:
\bea
\a\a R = - \frac {R_{i \bar \jmath m \bar n}\, F^i \bar F^{\bar \jmath} F^m \bar F^{\bar n}}{(F^k \bar F_k)^2} \,,\\
\a\a R' = - \frac {R_{i \bar \jmath m \bar n}\, F^i \bar F^{\bar \jmath} \hat F^m \hat {\bar F}^{\bar n}}
{(F^k \bar F_k)(\hat F^l \hat {\bar F}_l)} \,, \\
\a\a \hat R = - \frac {R_{i \bar \jmath m \bar n}\, \hat F^i \hat {\bar F}^{\bar \jmath} \hat F^m \hat {\bar F}^{\bar n}}
{(\hat F^k \hat {\bar F}_k)^2} \,, \\
\a\a M^2 = \frac {2 X_a^k \bar X_{bk}\, f^a_i \bar f^b_{\bar \jmath} \hat F^i \hat{\bar F}^{\bar \jmath}}{\hat F^l \hat{\bar F}_l} \,.
\eea
Note that compared to the treatment of $N=1$ theories with $F$ and $D$ breaking of section 3,  
the quantities $R$, $R'$, $\hat R$ and $M^2$ correspond to the quantities $R$, 
$S$, $T$ and $M^2$, whereas $F^i \bar F_i$ and $\hat F^i \hat {\bar F}_i$ correspond to 
$F^i \bar F_i$ and $\frac 12 D^a D_a$. Using the relation (\ref{relRTM}), we then see that the 
terms $R \, F^i \bar F_i$, $S \, D^a D_a$, $T \, (D^a D_a)^2/(4\,F^i \bar F_i)$ and $M^2 D^a D_a/F^i \bar F_i$ 
in eq.~(\ref{m+-FD}) become respectively $R \, F^i \bar F_i$, $2 R' \hat F^i \hat {\bar F}_i$, 
$R \, F^i \bar F_i - M^2 \hat F^i \hat {\bar F}_i/ F^j \bar F_j$ and $2 M^2 \hat F^i \hat {\bar F}_i/F^j \bar F_j$,
and there is some simplification in the masses of the first pair of sGoldstini. Concerning the 
second pair of sGoldstini, we now observe that they can actually be identified with particular 
real linear combinations of the would-be Goldstone modes 
$\sigma_a = \bar X_{ai} \phi^i + X_{a \bar \jmath} \phi^{\bar \jmath}$, and their conjugates
$\rho_a = i \bar X_{ai} \phi^i - i X_{a \bar \jmath} \phi^{\bar \jmath}$. Indeed, since 
$L^a X_a^i = 0$ and $L^a \bar X_a^{\bar \jmath} = - \s{\sqrt{2}} \hat {\bar F}^{\bar \jmath}$,
one has $\hat \varphi_+ = - \s{\sqrt{2}}\, {\rm Re}\,L^a \sigma_a = - \s{\sqrt{2}}\, {\rm Im}\,L^a \rho_a$ 
and $\hat \varphi_- = - \s{\sqrt{2}}\, {\rm Re}\,L^a \rho_a = \s{\sqrt{2}}\, {\rm Im}\,L^a \sigma_a$.
We moreover see that due to the fact that $X_a^i L^a= 0 $, we are in the situation where, as 
explained at the end of section 3, the Goldstone modes in the directions ${\rm Re}\,L^a$ and 
${\rm Im}\,L^a$ are linearly related to their conjugates in these directions. As a result, both 
$\hat \varphi_+$ and $\hat \varphi_-$ correspond to unphysical would-be Goldstone modes 
$\sigma_+$ and $\sigma_-$. This explains why they have vanishing masses, and also 
tells us that this information should be discarded. Taking the average of the first pair of 
sGoldstino masses, one is finally left with the following information:
\bea
\a\a m_{\varphi}^2 \equiv \s{\frac 12} \big(m_{\varphi_+}^2 \!+ m_{\varphi_-}^2\big) 
= 2\, R\, F^i \bar F_i + 2\, R'\, \hat F^i \hat{\bar F}_i + M^2 \frac {\hat F^i \hat{\bar F}_i}{F^j \bar F_j} \,.
\eea

Once again, the special form (\ref{RiemannSK}) taken by the Riemann tensor implies some relations among the 
quantities $R$, $R'$, $\hat R$ and $M^2$. More precisely, the stationarity condition implies that 
$C_{ijk} F^j F^k = \bar z^2 C_{ijk} \hat F^j \hat F^k + \s{\sqrt{2}} i \bar z^2 \bar X_{ai} f^a_j \hat F^j$ and leads to a relation 
between $R$, $\hat R$ and $M^2$, which is just eq.~(\ref{relRTM}) applied to the present case:
\be
R \, (F^i \bar F_i)^2 = \hat R \, (\hat F^i \hat {\bar F}_i)^2 + M^2 \hat F^i \hat {\bar F}_i \,.
\label{relRRhat2}
\ee
The expressions of the masses of the first pair of sGoldstini can then be recast in the following form:
\bea
\a\a m_{\varphi_\pm}^2 = 2\, R'\, \hat F^i \hat{\bar F}_i + 2\, \hat R\, \frac {(\hat F^i \hat{\bar F}_i)^2}{F^j \bar F_j}
+ 3\, M^2 \frac {\hat F^i \hat{\bar F}_i}{F^j \bar F_j} \pm \Delta \,. \label{m}
\eea
This finally yields:
\bea
\a\a m_{\varphi}^2 \equiv \s{\frac 12} \big(m_{\varphi_+}^2 \!+ m_{\varphi_-}^2\big) 
= 2\, R'\, \hat F^i \hat{\bar F}_i + 2\, \hat R\, \frac {(\hat F^i \hat{\bar F}_i)^2}{F^j \bar F_j} 
+ 3\, M^2 \frac {\hat F^i \hat{\bar F}_i}{F^j \bar F_j} \,.
\eea
This result corresponds to the information related to the first supersymmetry. 
We have seen that it can be obtained by simplifying the corresponding expression 
obtained in section 3 for $N=1$ theories with $F$ and $D$ breaking. There is instead 
no useful information related to the second supersymmetry, because the corresponding 
sGoldstini coincide with unphysical would-be Goldstone modes. 
Notice that in the limiting situations where $F^i \neq 0$ but $\hat F^i = 0$, 
the above positive-definite result for the average masses goes to zero. 
One is then back to a situation that is similar to the one arising in the 
Abelian case.

As before, one may now consider the more general situation with $e_a \neq 0$ and $\xi_a \neq 0$, 
where $F^i \neq 0$ and $\hat F^i \neq 0$. As for the Abelian case, we do not expect to get any new
information with this generalization, because all the $\xi_a$ can be set to zero through an overall 
$SU(2)$ transformation, provided the $N=2$ Fayet-Iliopoulos terms are aligned. It is nevertheless
instructive to work out the results also in this more general situation. In this case, we shall however 
not redo a detailed comparison with the $N=1$ perspective, and rather work out the results in a 
manifestly $SU(2)$ invariant way, in order to gain insight on how the information behaves 
under $SU(2)$. Using the notation (\ref{Pax}) and (\ref{Pa}), the four sGoldstino masses are found 
to be given by:
\bea
m_{\varphi_\pm}^2 \b=\b {\cal R}' \, P^{a0} P_a^0 + \hat{\cal R}\, \frac {(P^{a0} P_a^0)^2}{P^{bx} P_b^x} 
+ 3\, {\cal M}^2\, \frac {P^{a0} P_a^0}{P^{bx} P_b^x} \pm \Delta\,, \label{mnew} \\
m_{\hat \varphi_\pm}^2 \b=\b \frac {P^{a3}P_a^3}{P^{a3}P_a^3 + P^{a0} P_a^0}\, m_{\varphi_\pm}^2 \,, \label{mhatnew}
\eea
where
\bea
\a\a {\cal R}' = - \frac {R_{i \bar \jmath p \bar q} f^i_a \bar f^{\bar \jmath}_b f^p_c \bar f^{\bar q}_d\,
P^{ax} P^{bx} P^{c0} P^{d0}}{(P^{ey} P_e^y)(P^{f0} P_f^0)} \,, \\
\a\a \hat {\cal R} = - \frac {R_{i \bar \jmath p \bar q} f^i_a \bar f^{\bar \jmath}_b f^p_c \bar f^{\bar q}_d\,
P^{a0} P^{b0} P^{c0} P^{d0}}{(P^{e0} P_e^0)^2} \,, \\
\a\a {\cal M}^2 = \frac {2\,X_a^i \bar X_{bi} \, P^{a0} P^{b0}}{P^{c0} P_c^0} \,.
\eea
We see that (\ref{mnew}) is simply the $SU(2)$ invariant completion of (\ref{m}), and therefore represents 
the correct generalization of the information. On the other hand, (\ref{mhatnew}) is not $SU(2)$ invariant
and does not represent any additional information. The reason is that when $\xi_a \neq 0$, the 
two directions $F^i$ and $\hat F^i$ are no-longer orthogonal. The most appropriate way to proceed is 
then to subtract from $\hat F^i$ its projection $F_{\t{\parallel}}^i$ along $F^i$, and look at the direction 
$\hat F^i_{\t{\perp}}$. But this direction is nothing but the complex would-be Goldstone 
direction $X_a^i \bar L^a$, corresponding to the unphysical modes 
$\sigma_+ = - \s{\sqrt{2}}\, {\rm Re}\,L^a \sigma_a = - \s{\sqrt{2}}\, {\rm Im}\,L^a \rho_a$ 
and $\sigma_- = - \s{\sqrt{2}}\, {\rm Re}\,L^a \rho_a = \s{\sqrt{2}}\, {\rm Im}\,L^a \sigma_a$, which lead to 
vanishing masses. This shows that (\ref{mhatnew}) represents in fact 
the same information as (\ref{mnew}), but diluted along an unphysical direction.
So once again the only useful information comes 
from the first pair of sGoldstini, and reads:
\bea
\a\a m_{\varphi}^2 \equiv \s{\frac 12} \big(m_{\varphi_+}^2 \!+ m_{\varphi_-}^2\big) =
{\cal R}' \, P^{a0} P_a^0 + \hat{\cal R}\, \frac {(P^{a0} P_a^0)^2}{P^{bx} P_b^x} 
+ 3\, {\cal M}^2\, \frac {P^{a0} P_a^0}{P^{bx} P_b^x} \,. \label{mNA1}
\eea

One may wonder whether it is possible to get this $SU(2)$-invariant information in a more transparent 
way, by somehow reorganizing the four sGoldstini according to their $SU(2)$ transformation properties, 
as in the case of the hyper multiplets. To answer this question, notice first that in this case, contrarily to the case 
involving only hypers, the Lagrangian is not $SU(2)$-invariant, unless one promotes the Fayet-Iliopoulos 
constants $P_a^x$ to triplet spurions. The transformation properties of the sGoldstini are then determined 
by the dependence of the Goldstino directions on the singlets $P_a^0$ and the triplets $P_a^x$. Notice in 
this respect that we have defined the two Goldstino directions in terms of 
$F^i \propto \bar f^{ia} (P_a^1 - i P_a^2)$ and $\hat F^i \propto \bar f^{ia} (P_a^0 + P_a^3)$. But one could 
have equivalently used also the other two quantities $f^i_a D^a \propto \bar f^{ia} (P_a^0 - P_a^3)$ 
and $f^i_a \hat D^a \propto \bar f^{ia} (P_a^1 + i P_a^2)$; these would have given the same information
in the above analysis, as a consequence of the alignment of the triplets $P_a^x$ and the relation of the singlets 
$P_a^0$ to would-be Goldstone modes. Then, considering all these four complex directions on equal footing 
one might equally well switch to the linear combinations $\bar f^{ia} P_a^0$ and $\bar f^{ia} P_a^x$, which are 
clearly a singlet and a triplet of $SU(2)$. Notice however that due to the alignment condition $P_a^x = p_a P^x$, 
the latter three vectors differ only by their normalization, and define thus the same direction. In this way one 
recovers just two independent complex directions, which are both $SU(2)$ invariant, and the masses of the 
corresponding pairs of real sGoldstini are respectively given by $0\pm 0$ and $m_\varphi^2 \pm \Delta$,
with $m_\varphi^2$ given by eq.~(\ref{mNA1}).

The above result is new. It shows that the situation improves when generalizing the gauging from 
Abelian to non-Abelian. Tachyons do no longer necessarily appear, because those states that were 
giving rise to them in the Abelian case receive an additional positive definite contribution to their mass 
in the non-Abelian case. Note however that when $P_a^x = 0$ one gets $P_a^0 = 0$ at stationary 
points, by the reasoning after (\ref{relDFF}). It is thus necessary to switch on at least some of the 
$P_a^x$ to achieve metastability.
Another case where the result (\ref{mNA1}) vanishes identically is when the prepotential is quadratic, 
since in that case ${\cal R'}$ and $\hat {\cal R}$ vanish due to the vanishing of the curvature and ${\cal M}^2$ 
vanishes due to eq.~(\ref{DFFvec}) contracted with $D^a$ and the constancy of the gauge
kinetic function. This is compatible with what happens in the rigid limit 
of the examples constructed in \cite{FTV}, where for $M_{\rm P} \to \infty$ the 
geometry becomes flat and the scalar masses tend to zero.

We expect that to obtain the generalization of this result to supergravity, one should proceed exactly 
along the same lines and compute the average mass of the first pair of sGoldstini. But as usual, the 
supergravity result can differ from the rigid one derived here only by quantitative effects, suppressed 
by inverse powers of the Planck scale. One should then be left with some freedom to keep the value 
of the average mass positive also in the presence of gravity. Concerning the second pair of sGoldstini, 
we believe that they are again associated to two would-be Goldstone modes, and do therefore not yield 
any further information. Indeed, the relevant direction in group space is changed from $L^a$ to $L^A$, 
with $A=0,a$ and involves now also the graviphoton direction, but the crucial property $X_a^i L^a = 0$
simply generalizes to $X_A^i L^A = 0$. As a result, it remains true also in supergravity that these two 
modes are both massless but unphysical. We have verified this statement in the explicit examples 
constructed in \cite{FTV}, where there is always a pair of would-be Goldstone modes forming a complex 
scalar field.

In this case too it is unclear to what extent the necessary condition for metastability could
be made sufficient by allowing a tuning. Indeed, for a given geometry associated to $K$ 
the only things one may change are the Killing potentials defining the gauge symmetries. 
But these are not arbitrary functions, and can therefore be adjusted only in a limited way.

\section{N=2 models with hyper and vector multiplets}
\setcounter{equation}{0}

Let us finally consider the most general case of $N=2$ theories with $n_{\cal H}$ hyper multiplets ${\cal H}^k$
and $n_{\cal V}$ vector multiplets ${\cal V}^{a}$. This is a particular case of $N=1$ theory containing
$n_{\rm C} = 2 n_{\cal H} + n_{\cal V}$ chiral multiplets $Q^u$ and $\Phi^i$ plus $n_{\rm V} = n_{\cal V}$ 
vector multiplets $V^a$. The most general two-derivative Lagrangian is specified by a real K\"ahler potential 
$K$, a holomorphic superpotential $W$, a holomorphic gauge kinetic function $f_{ab}$, some triholomorphic
and holomorphic Killing vectors $X_a^u$ and $X_a^i$, and some real Fayet-Iliopoulos constants 
$\xi_a$, all subject to strong restrictions required for the existence of a second supersymmetry. We shall not derive 
in full detail these restrictions, because they emerge essentially in the same way as in the cases involving only 
hyper and vector multiplets, discussed in sections 4 and 5. Moreover we shall restrict from the beginning 
to theories where the superpotential involves only an electric term and no magnetic 
term. In $N=1$ superspace, the Lagrangian is then found to take the following form:
\bea
\mathcal{L} \b=\b \int \! d^4 \theta\, \Big[K^H (Q,\bar Q,V)  + K^V(\Phi,\bar \Phi,V) + \xi_a V^a \Big] \\
\b\;\b + \int \! d^2 \theta\, \Big[s P(Q)  + \s{\sqrt{2}}\, e_a L^a(\Phi) + \s{\sqrt{2}}i \, P_a(Q) L^a(\Phi) 
- \frac i4\, M_{ab} (\Phi) \,W^{a \alpha} W_\alpha^b \Big] + {\rm h.c.} \,. \nn 
\eea
Besides the normal coupling between hyper and vector multiplets, which involves the real Killing potentials $- \frac 12 K_a^H$
associated to the Killing vectors $X_a^u$, there is also an additional coupling which involves the holomorphic Killing 
potentials $P_a$ admitted by the $X_a^u$ due to the fact that they are triholomorphic. These extra couplings are required 
by the second supersymmetry, and generalize the well-known couplings arising already in the minimal theory
based on a flat geometry between the pair of chiral multiplets forming each hyper multiplet and the adjoint scalar 
contained in each vector multiplet. The self-interaction of hyper multiplets, which represents the generalization 
of the hyper multiplet mass terms in the flat case, are again described by a triholomorphic Killing vector 
$X^u = \s{\sqrt{2}}i \bar s\, X^u_0 L^0$, and the associated holomorphic Killing potential $P = \s{\sqrt{2}}i \bar s\, P_0 L^0$.

The above Lagrangian is invariant under a second supersymmetry, which acts on the $N=1$ superfields
in the following way:
\bea
\a\a \hat \delta Q^u = -\s{\frac 12} \bar \Omega^{uv} \bar D^2 \big[\big(K_v(Q,\bar Q) + 2i \bar X_{av} (Q,\bar Q) V^a + {\cal O}(V^2) \big)
(\hat \epsilon \theta + \hat {\bar \epsilon} \bar \theta) \big] \nn \\
\a\a \hspace{34pt} -\, 2 i \big[(s + \bar s)\, X^u(Q) + \s{\sqrt{2}}i\, X^u_a (Q,\bar Q) L^a(\Phi) \big]
\hat \epsilon \theta \,, \label{deltahatg1} \\[1mm]
\a\a \hat \delta \Phi^i = \s{\sqrt{2}} i f_a^i(\Phi) \hat \epsilon W^a \,, \label{deltahatg2} \\[1.5mm]
\a\a \hat \delta V^a = - \s{\sqrt{2}} i \big(\bar L^a (\bar \Phi) 
- i \, f_{bc}^{\;\;\,a} \bar L^b(\bar \Phi) V^c + {\cal O}(V^2) \big) \hat \epsilon \theta + {\rm h.c.} \,. \label{deltahatg3}
\eea
The full $N=2$ supersymmetry algebra closes only on-shell, by using the equations of motion of the superfields $Q^u$ describing the 
hyper multiplets, and there is a central charged acting on the latter:
\bea
\a\a \delta_{\rm c} Q^u = \alpha X^u(Q) \,, \\
\a\a \delta_{\rm c} \Phi^i = 0 \,, \\
\a\a \delta_{\rm c} V^a = 0 \,.
\eea

One may again use alternative forms of the supersymmetry transformations, which are equivalent 
on-shell for the $Q^u$. For instance, one may add to (\ref{deltahatg1}) the 
trivial transformation $\delta_{\rm t} Q^u = \frac 12 \Omega^{u v} [\bar D^2 (K_v + 2i \bar X_{av} V^a + {\cal O}(V^2)) 
- 4 s P_v - 4 \s{\sqrt{2}}i P_{av} L^a] \hat \epsilon \theta$, which is a symmetry of the on-shell theory since 
the parenthesis is proportional to the equations of motion of $Q^u$. 
This gives $\hat \delta Q^u = - \frac 12 \bar \Omega^{uv} \bar D^2[(K_v + 2i \bar X_{av} V^a + {\cal O}(V^2)) \hat {\bar \epsilon} \bar \theta] 
- 2 i \bar s X^u \hat \epsilon \theta$.

The gauge transformations are defined by the triholomorphic Killing vector $X_a^u$ for $Q^u$, and 
take the same fixed form as before for $\Phi^i$ and $V^a$, corresponding to the adjoint representation:
\bea
\a\a \delta_{\rm g} Q^u = \Lambda^a X_a^u \,, \\[1mm]
\a\a \delta_{\rm g} \Phi^i= f^i_a f_{bc}^{\;\;\; a} \Lambda^b L^c \,, \\
\a\a \delta_{\rm g} V^a = - \s{\frac i2} \big(\Lambda^a - \bar \Lambda^a) 
+ \s{\frac 12} f_{bc}^{\;\;\;a} \big(\Lambda^b + \bar \Lambda^b) V^c + {\cal O}(V^2) \,.
\eea

The Killing vectors are related to the Killing potentials in the usual way, both in the hyper and in the vector 
multiplet sectors:
\be
X_a^i = \s{\frac i2} g^{i \bar \jmath} \nabla_{\bar \jmath} K_a^V \,,\;\;
X_a^u = \s{\frac i2} g^{u \bar v} \nabla_{\bar v} K_a^H \,.
\ee
The equivariance conditions following from the fact that these Killing vectors $X_a^i$ and $X_a^u$ are holomorphic 
take the usual form:
\be
g_{i \bar \jmath} X_{\t{[}a}^i \bar X_{b\t{]}}^{\bar \jmath} = \s{\frac i4} f_{ab}^{\;\;\; c} K_c^V \,,\;\;
g_{u \bar v} X_{\t{[}a}^u \bar X_{b\t{]}}^{\bar v} = \s{\frac i4} f_{ab}^{\;\;\; c} K_c^H \,.\;\;
\label{equi1}
\ee
In addition, there is an other equivariance condition emerging in the hyper multiplet sector,
due to the fact that $X_a^u$ is actually triholomorphic. More precisely, exploiting the fact that 
it is also holomorphic with respect to the two extra complex structures yields the following 
extra complex condition, involving the holomorphic Killing potential $P_c$:
\be
\Omega_{uv} X_a^u X_b^v = i f_{ab}^{\;\;\; c} P_c \,.
\label{equi2}
\ee
We see that this condition is actually crucial to guarantee the gauge invariance of the 
term in the superpotential that mixes hyper and vector multiplets. Finally, global central 
charge invariance of the minimal gauge coupling $K_a V^a$ and gauge invariance
of the superpotential $P = \s{\sqrt{2}}i \bar s P_0 L^0$ for hyper multiplets impose two further constraints, 
one real and one complex, which read:
\be
g_{u \bar v} X_{\t{[}0}^u \bar X_{a\t{]}}^{\bar v} = 0 \,,\;\;  \Omega_{uv} X_0^u X_a^v = 0 \,.
\label{equi3}
\ee
These conditions ensure the compatibility between the local gauge symmetry 
and the global central charge symmetry, which are independent.

In the Wess-Zumino gauge, the action can be expanded at quadratic order in the vector superfields
and simplifies to the following expression:
\bea
\mathcal{L} \b=\b \int \! d^4 \theta\, \Big[K^H(Q,\bar Q)  + K^V(\Phi,\bar \Phi) + \big(K^H_a(Q,\bar Q) + K^V_a(\Phi, \bar \Phi) + \xi_a \big) V^a \nn \\
\b\;\b \hspace{33pt} +\, 2\, \big(g_{u \bar v} (Q,\bar Q) X_a^u (Q) \bar X_b^{\bar v} (\bar Q) + 
g_{i \bar \jmath} (\Phi,\bar \Phi) X_a^i (\Phi) \bar X_b^{\bar \jmath} (\bar \Phi) \big) V^a V^b \Big] \nn \\
\b\;\b + \int \! d^2 \theta\, \Big[s P(Q) + \s{\sqrt{2}} \big(i P_a(Q) + e_a \big) L^a(\Phi) 
- \s{\frac i4}\, M_{ab} (\Phi) \,W^{a \alpha} W_\alpha^b \Big] + {\rm h.c.} \,.
\eea
We see now that much as the real constants $\xi_a$ correspond to the ambiguity in the real Killing potentials 
$K_a^H$, the complex constants $e_a$ correspond to the ambiguity in the holomorphic Killing 
potentials $P_a$, for Abelian factors. Moreover, one may now verify more explicitly the invariance of 
the couplings between hyper and vector multiplets, by keeping terms with up to one vector multiplet in 
eqs.~(\ref{deltahatg1})-(\ref{deltahatg3}). In components, one finds:
\bea
\mathcal{L} \b=\b -g_{u \bar v} \, D_\mu q^u D^\mu \bar q^{\bar v} - g_{i \bar \jmath} \, D_\mu \phi^i D^\mu \bar \phi^{\bar \jmath}
- \s{\frac 14} h_{ab} \, F^a_{\mu\nu} F^{b\mu\nu} + \s{\frac 14} k_{ab} \, F^a_{\mu\nu}\tilde{F}^{b\mu\nu} \nn \\
\b\;\b -i g_{u \bar v} \, \chi^u \big(D\!\!\!\!/\,\bar \chi^{\bar v} + \Gamma^{\bar v}_{\bar s \bar t} \, D\!\!\!\!/\, \bar q^{\bar s} \bar \chi^{\bar t} \big)
-ig_{i \bar \jmath} \, \psi^i \big(D\!\!\!\!/\,\bar \psi^{\bar \jmath} 
+ \Gamma^{\bar \jmath}_{\bar m \bar n} \, D\!\!\!\!/\, \bar \phi^{\bar m} \bar \psi^{\bar n} \big)  
- \s{\frac i2} h_{ab} \, \lambda^{a} D\!\!\!\!/\, \bar \lambda^b + {\rm h.c.} \nn \\
\b\;\b - \s{\frac i{\!\sqrt{2}}} C_{ijk} f^j_a f^k_b \, \lambda^a \sigma^{\mu \nu} \psi^i F^{b}_{\mu\nu} + {\rm h.c.} 
- V_{\rm S} - V_{\rm F} \,, 
\eea
where:
\bea
V_{\rm S} \b=\b g_{u \bar v} (s X^u \!+ \s{\sqrt{2}}i\, X^u_a L^a) (\bar s \bar X^{\bar v} \! - \s{\sqrt{2}}i\, \bar X_a^{\bar v} \bar L^a)
+ 2 \, h^{ab} \big(P_a -i e_a \big) \big(\bar P_b + i \bar e_b\big) \nn \\
\b\;\b + \s{\frac 18} \, h^{ab}(K_a^H \!+ K_a^V \!+ \xi_a)(K_b^H \!+ K_b^V \!+ \xi_b) \,, \label{VShypvec} \\[-0.5mm]
V_{\rm F} \b=\b \s{\frac 12} \Big[i\, \Omega_{uw} \nabla_{v} (s X^w \!+ \s{\sqrt{2}}i\, X_a^w L^a ) \, \chi^u \chi^v
- \s{\sqrt{8}}\, \Omega_{uv} X_a^v f_i^a \chi^u \psi^i + \s{\sqrt{8}}\, \bar X_{au} \chi^u \lambda^a \nn \\[0.5mm]
\b\;\b \hspace{14pt} -\, \s{\sqrt{2}}\, C_{ijk} \bar f^{ka} \big((P_a - i e_a) \,\psi^i \psi^j 
- (\bar P_a + i \bar e_a)  f^i_b f^j_c \lambda^b \lambda^c \big) \nn \\[1mm]
\b\;\b  \hspace{14pt} +\, \s{\sqrt{8}} \big(\bar X_{ai} + \s{\frac 14} C_{ijk} f^j_a \bar f^{kb} (K_b^H \!+ K_b^V \!+ \xi_b) \big)
\psi^i \lambda^a \Big] + {\rm h.c.} \nn \\
\b\;\b \!- \s{\frac 14} R_{u \bar v s \bar t} \, \chi^u \chi^s \bar \chi^{\bar v} \bar \chi^{\bar t}
- \! \s{\frac 14} R_{i \bar \jmath k \bar l} \big(\psi^i \psi^k \bar \psi^{\bar \jmath} \bar \psi^{\bar l}
\!+\! f^i_a f^k_b \bar f^{\bar \jmath}_c \bar f^{\bar l}_d \, \lambda^a \lambda^b \bar \lambda^c \bar \lambda^d 
\!+\! 2 f^k_a \bar f^{\bar l}_b\, \psi^i \lambda^a \bar \psi^{\bar \jmath} \bar \lambda^b \big) \nn \\
\b\;\b \!+ \s{\frac 14} \Big[\big(i \nabla_i C_{j k l} + 2\, C_{ikm} C_{jln} f^m_c \bar f^{nc} \big) f^k_a f^l_b\, \psi^i \psi^j \lambda^a \lambda^b \nn \\[0mm]
\b\;\b \hspace{20pt} +\, C_{ikm} C_{jln} f^m_c \bar f^{nc} f^k_a f^l_b \, \psi^i \lambda^a \psi^j \lambda^b 
\Big]+ {\rm h.c.} \,.
\eea

To determine the first supersymmetry transformation laws in components, one has as usual to take into 
account the need for a compensating gauge transformation to stay in the Wess-Zumino gauge, with 
parameter $\Lambda^{a} = 2 i \theta \sigma^\mu \bar \epsilon A^a_\mu + 2 \theta^2 \bar \epsilon \bar \lambda^a$. 
The additional gauge transformation turns the ordinary derivatives appearing in $\delta \chi^u$ and $\delta \psi^i$ 
into gauge-covariant derivatives, and one finds
\bea
\a\a \delta q^u = \s{\sqrt{2}}\, \epsilon \,\chi^u \,, \\
\a\a \delta \chi^u = \s{\sqrt{2}}\, \epsilon \,F^u + \s{\sqrt{2}}i \,D\!\!\!\!/\, q^u \bar \epsilon \,, \label{dchig} \\
\a\a \delta \phi^i = \s{\sqrt{2}}\, \epsilon\, \psi^i \,, \\
\a\a \delta \psi^i = \s{\sqrt{2}}\, \epsilon\, F^i + \s{\sqrt{2}}i\,D\!\!\!\!/\, \phi^i \, \bar \epsilon \,, \label{dpsig} \\
\a\a \delta A_\mu^a = i \epsilon\, \sigma_\mu \bar \lambda^a - i \lambda^a \sigma_\mu \bar \epsilon \,, \\
\a\a \delta \lambda^a = i \epsilon\, D^a + \sigma^{\mu \nu} \epsilon\, F_{\mu \nu}^a \,, \label{dlambdag}
\eea
The auxiliary fields $F^u$, $F^i$ and $D^a$ are given by
\bea
\a\a F^u = i \bar \Omega^u_{\;\;\bar v} (\bar s\, \bar X^{\bar v} \!-i \s{\sqrt{2}} \bar X^{\bar v}_a \bar L^a) + \s{\frac 12} \Gamma^u_{st}\, \chi^s \chi^t \,, \\
\a\a F^i = \s{\sqrt{2}}i\, \bar f^{ia} (\bar P_a + i \bar e_a) + \s{\frac 12} \Gamma^i_{mn} \, \psi^m \psi^n 
+ \s{\frac i2} \bar C^i_{\;\,\bar m \bar n} \bar f^{\bar m}_a \bar f^{\bar n}_b \, \bar \lambda^a \bar \lambda^b \,, \\
\a\a D^a = - \s{\frac 12} h^{ab} (K_b^H \! + K_b^V \!+ \xi_b) - \s{\frac 1{\sqrt{2}}}\, C_{i j k} \bar f^{ja} f^k_b \, \psi^i \lambda^b + {\rm h.c.} \,.
\eea
To determine the second supersymmetry transformation laws, one has to similarly supplement the transformations 
(\ref{deltahatg1})--(\ref{deltahatg3}) with a compensating gauge transformation to stay in the Wess-Zumino gauge,
with parameter $\hat \Lambda^{a} = - \s{\sqrt{8}} \theta \hat \epsilon \bar L^a - 2 \theta^2 \! f_i^a \hat \epsilon \psi^i$.
The additional gauge transformation shifts the $\bar F^{\bar v}$ auxiliary field appearing in 
$\hat \delta \chi^u$ by $- \s{\sqrt{2}}i P_a^{\bar v} \bar L^a$ and the $D^a$ auxiliary field appearing in 
$\hat \delta \psi^i$ by $h^{ab} K_b$, and one finds
\bea
\a\a \hat \delta q^u = - \s{\sqrt{2}}\, \bar \Omega^u_{\;\;\bar v} \, \hat {\bar \epsilon} \,\bar \chi^v \,, \\
\a\a \hat \delta \chi^u = \s{\sqrt{2}}\, \hat \epsilon \, \hat F^u
+ \s{\sqrt{2}}\,\Gamma^u_{st} \bar \Omega^s_{\;\; \bar v}  \, \hat {\bar \epsilon} \, \bar \chi^{\bar v} \chi^t 
+ \s{\sqrt{2}}i\,\bar \Omega^u_{\;\;\bar v} D\!\!\!\!/\, \bar q^{\bar v} \hat {\bar \epsilon} \label{dhatchig}\,, \\
\a\a \hat \delta \phi^i = \s{\sqrt{2}}\, \hat \epsilon\, f^i_a \lambda^a \,, \label{dhatphig} \\
\a\a \hat \delta \psi^i = \s{\sqrt{2}} \hat \epsilon\, \hat F^i+ \s{\sqrt{2}}\, \partial_j f_a^i \psi^j (\hat \epsilon \lambda^a) 
+ \sigma^{\mu \nu} \hat \epsilon\, f^i_a F_{\mu \nu}^a \,, \label{dhatpsig} \\
\a\a \hat \delta A_\mu^a = - i\, \hat \epsilon\, \sigma_\mu \bar f^a_{\bar \imath} \bar \psi^{\bar \imath} 
+ i f^a_i \psi^i \sigma_\mu\, \hat{\bar \epsilon} \,, \\[0mm]
\a\a \hat \delta \lambda^a = i \hat \epsilon\, \hat D^a 
+ \s{\sqrt{2}}i\,f^a_i D\!\!\!\!/\, \phi^i \, \hat {\bar \epsilon} \,. \label{dhatlambdag}
\eea
The quantities $\hat F^u$, $\hat F^i$ and $\hat D^a$ are found to be given by
\bea
\a\a \hat F^u = \bar \Omega^u_{\;\; \bar v} \big(\bar F^{\bar v} \!+ (s + \bar s) P^{\bar v} \!+\! \s{\sqrt{2}}i (L^a \!-\! \bar L^a) P_a^{\bar v} 
\!- \s{\frac 12}\,\Gamma^{\bar v}_{\bar s \bar t} \,\bar \chi^{\bar s} \bar \chi^{\bar t} \big)
= - i \bar s X^u \!-\! \s{\sqrt{2}} X_a^u \bar L^a , \\[0.5mm]
\a\a \hat F^i = \s{\frac i{\sqrt{2}}} f^i_a \big(D^a + h^{ab} K_b^V\big)
= \s{\frac i{\sqrt{8}}} \bar f^{ia} (K_a^V \!- K_a^H \!- \xi_a)  + \text{ferm.}  \,, \\
\a\a \hat D^a = - \s{\sqrt{2}} i \big( \bar f^a_{\bar \imath} \bar F^{\bar \imath} 
- \s{\frac 12} \,\partial_{\bar \imath} f_{\bar \jmath}^a \bar \psi^{\bar \imath} \bar \psi^{\bar \jmath} \big) 
= - 2\, h^{ab} (P_b - i e_b) + \text{ferm.} \,.
\eea

The extension to supergravity can be found in \cite{dWvP,dWLvP,DFF,ABCDFFM}. It turns again out that 
models of the above type can be consistently coupled to gravity only if the coefficients of the 
Fayet-Iliopoulos terms and the part of the superpotential linear in the sections satisfy some restrictions. 
The main new feature is that there appears a non-trivial $SU(2)$ bundle over the hyper multiplet 
scalar manifold and a non-trivial $U(1)$ bundle over the vector multiplet scalar manifold, 
with curvatures proportional to $M_{\rm P}^{-2}$, and the full scalar manifold becomes the 
product of a Quaternionic-K\"ahler manifold and a Special-K\"ahler-Hodge manifold. 
To spell out more precisely the restrictions that need to be imposed on the $N=2$
Fayet-Iliopoulos terms, we proceed as before and relabel the various Killing potentials 
in a more appropriate way, by defining a triplet of new potentials $P_a^x$ as follows:
\be
P_a^1 = - 2\, {\rm Im} (P_a \!- i e_a) \,,\;\;
P_a^2 = 2\, {\rm Re} (P_a \!- i e_a) \,,\;\; 
P_a^3 = \s{\frac 12} \big(K_a^H \!+ \xi_a\big) \,,\;\; 
\ee
One may also introduce as before the notation
\be
P_a^0 = - \s{\frac 12} K_a^V \,.
\ee
The triplet of functions $P_a^x$ must satisfy a non-trivial equivariance condition, and 
are thus constrained. As before, there is a non-trivial effect coming from the curvature of the 
$SU(2)$, which is of order $M_{\rm P}^{-2}$ and is thus a genuine supergravity effect. 
For Abelian factors under which no hyper multiplet is charged, however, this is the only term 
that arises, and one then obtains a constraint that is independent of $M_{\rm P}^{-2}$, and 
survives thus in the rigid limit. This constraint takes the form (\ref{align}), whose solution is (\ref{solalign}).
For non-Abelian factors, on the other hand, the gravitational deformation of the 
equivariance condition is smooth and can be safely discarded in the rigid limit. 
One is then left with the equivariance conditions (\ref{equi1})--(\ref{equi3}). 
Notice finally that the superpotential does no longer display the special property 
(\ref{Wspec}), because it now involves also the non-Abelian sections. 

Notice finally that it is possible to reshuffle the scalar potential 
(\ref{VShypvec}) by proceeding in the following way, with manipulations that are similar to 
those used in \cite{ADF1,ADF2} to discuss truncations of $N=2$ 
to $N=1$ supergravity theories. From now on we set again $s=i$ for simplicity.
For the $F$-term part from the hyper multiplets, we start by rewriting it as 
$2 g_{u \bar v} X_A^u \bar X_B^{\bar v} L^A \bar L^B$, with a new index $A=0,a$ 
comprising both the Killing vector defining the central charge global symmetry and 
the Killing vectors defining the gauge symmetry. Since this ranges 
over at least two values, both the symmetric and the antisymmetric parts 
of $g_{u \bar v} X_A^u \bar X_B^{\bar v}$ contribute. For the symmetric part, we may 
proceed as in the case with only hypers, and switch to general real coordinates by rewriting
$g_{u \bar v} X_{\t{(}A}^u \bar X_{B\t{)}}^{\bar v} = \frac 12 g_{UV} X_A^U X_B^V$, which gives
$2 g_{u \bar v} X_{\t{(}A}^u \bar X_{B\t{)}}^{\bar v} L^A \bar L^B = g_{UV} X_A^U X_B^V L^A \bar L^B$.
For the antisymmetric part, the equivariance relations (\ref{equi1}) and (\ref{equi3}) imply
$g_{u \bar v} X_{\t{[}a}^u \bar X_{b\t{]}}^{\bar v} = \frac i4 f_{ab}^{\;\;\; c} K_c^H$ and 
$g_{u \bar v} X_{\t{[}0}^u \bar X_{a\t{]}}^{\bar v} = 0$, and therefore 
$2 g_{u \bar v} X_{\t{[}A}^u \bar X_{B\t{]}}^{\bar v} L^A \bar L^B = \frac i2 f_{ab}^{\;\;\; c} K_c^H L^a \bar L^b
= - \frac 14 h^{ab} K_a^V K_b^H$. Putting everything together, we see that the $F$-term 
part coming from the hyper multiplets finally gives $2 g_{u \bar v} X_A^u \bar X_B^{\bar v} L^A \bar L^B = 
g_{UV} X_A^U X_B^V L^A \bar L^B - \frac 14h^{ab} K_a^V K_b^H$.
For the $F$-term part of the vectors, we get instead
$2 h^{ab} \,(P_a -i e_a) (\bar P_b +i \bar e_b) = \frac 12 h^{ab} (P_a^1 P_b^1 + P_a^2 P_b^2)$.
Finally for the $D$-term part it is convenient to consider separately the three types of terms that arise respectively 
from hyper multiplets, from vector multiplets and from their interference. For the vector multiplet part, we have 
as before $\frac 18 h^{ab} K_a^V \bar K_b^V = \frac 12 g_{i \bar \jmath} X_a^i \bar L^a \bar X_b^j L^b = \frac 12 h^{ab} P_a^0 P_b^0$. 
For the hyper multiplet part, we get $\frac 18 h^{ab} (K_a^H \!+ \xi_a) (K_b^H \!+ \xi_b)  = \frac 12 h^{ab} P_a^3 P_b^3$. 
Finally, for the mixed part we get $\frac 14 h^{ab} K_a^V (K_b^H \!+ \xi_b) = \frac 14 h^{ab} K_a^V K_b^H$. Collecting 
the above results for the three terms in (\ref{VShypvec}), we see that the interference terms involving $h^{ab} K_a^V K_b^H$ 
cancel out, and the scalar potential can finally be rewritten in the following form:
\bea
V_{\rm S} \b=\b g_{UV} X_A^U \bar L^A X_B^V L^B + \s{\frac 12} g_{i \bar \jmath} X_a^i \bar L^a \bar X_b L^b 
+ \s{\frac 12} h^{ab} P_a^x P_b^x \nn \\
\b=\b g_{UV} X_A^U \bar L^A X_B^V L^B + \s{\frac 12} h^{ab} \big(P_a^0 P_b^0 + P_a^x P_b^x \big) \,.
\eea
Notice also that the equivariance conditions (\ref{equi1})--(\ref{equi3}) can be rewritten in the following more 
compact form:
\bea
i g_{i \bar \jmath} X_{\t{[}a}^i \bar X_{b\t{]}}^{\bar \jmath} = \s{\frac 12} f_{ab}^{\;\;\; c} P_c^0 \,,\;\;
J^x_{UV} X_{\t{[}A}^U X_{B\t{]}}^V = f_{AB}^{\;\;\; C} P_C^x \,.
\eea
Here $f_{AB}^{\;\;\;C}$ denote the structure constants of the group $G \times U(1)$ defined by the gauge 
group $G$ and the $U(1)$ central charge symmetry, such that $f_{ab}^{\;\;\;c}$ are the structure constants 
of the gauge group and $f_{0a}^{\;\;\; b} = f_{ab}^{\;\;\; 0} = 0$. This rewriting reflects once again the fact that the 
superpotential for the hyper multiplets comes in supergravity from a gauging of the central charge 
by the graviphoton $A_\mu^0$, which is then treated on equal footing with the other gauge fields
$A_\mu^a$. It also shows that in order for the graviphoton gauging to leave a remnant in the rigid 
limit, it must be associated to a factorized $U(1)$.

\subsection{Supertrace}

At a generic point in the scalar field space and for vanishing fermions and vector fields, the auxiliary fields 
simplify to
\bea
\a\a F^u = \s{\sqrt{2}}\,\bar \Omega^u_{\;\;\bar v} \bar X^{\bar v}_A \bar L^A \,, \\[0.5mm]
\a\a F^i = \s{\sqrt{2}}i\,\bar f^{ia} (\bar P_a + i \bar e_a) \,,\\
\a\a D^a = - \s{\frac 12} h^{ab} (K_a^H \!+ K_a^V \!+ \xi_b) \,.
\eea
The corresponding hatted quantities similarly simplify to
\bea
\a\a \hat F^u = \bar \Omega^u_{\;\; \bar v} \big(\bar F^{\bar v} \!+\! \s{\sqrt{2}}i (L^A \!-\! \bar L^A) P_A^{\bar v} \big)
= - \s{\sqrt{2}}\, X_A^u \bar L^A \,, \label{Fhag} \\[0.5mm]
\a\a \hat F^i = \s{\frac i{\sqrt{2}}} f^i_a \big(D^a + h^{ab} K_b^V\big) 
= \s{\frac i{\sqrt{8}}} \bar f^{ia} (K_a^V \!- K_a^H \!- \xi_a) \,, \\
\a\a \hat D^a = - \s{\sqrt{2}} i\, \bar f^a_{\bar \imath} \bar F^{\bar \imath} =  - 2\, h^{ab} (P_b - i e_b) \,.
\eea
The mass matrix of the scalar fields is given by
\bea
\a\a (m_0^2)_{u \bar v} = 2\, \nabla_u X^w_A L^A \nabla_{\bar v} \bar X_{Bw} \bar L^B
+ 2\, \Omega_{us} \bar \Omega_{\bar v \bar t} h^{ab} X_a^s \bar X_b^{\bar t} \nn \\
\a\a  \hspace{45pt} -\, R_{u \bar v s \bar t} \, F^s \bar F^{\bar t} + h^{ab} \bar X_{au} X_{b \bar v} 
+ \s{\frac i2} \nabla_u X_{a \bar v} D^a + {\rm h.c.} \,, \\
\a\a (m_0^2)_{i \bar \jmath} = - R_{i \bar \jmath k \bar l}\, \big(F^k \bar F^{\bar l} + f_a^k \bar f^a_{\bar m} \bar f_b^{\bar l} f^b_{n} F^n \bar F^{\bar m} 
+ f^k_a \bar f^{\bar l}_b\, D^a D^b \big) 
+ 2\, f_i^a \bar f_{\bar \jmath}^b X_a^s \bar X_{bs} \nn \\[0mm]
\a\a \hspace{45pt} +\, h^{ab} \bar X_{a i} X_{b \bar \jmath} 
+ \s{\frac i2} \big(\nabla_i X_{a \bar \jmath} - 2 h^{bc} h_{abi} X_{c \bar \jmath} \big)\,D^a + {\rm h.c.} \,, \\
\a\a (m_0^2)_{u \bar \imath} = 2\, \nabla_u X_{A \bar v} L^A \bar X_b^{\bar v} \bar f_{\bar \imath}^b
- \s{\sqrt{2}}i\, \Omega_{u s} X_a^s \bar C_{\bar \imath \bar \jmath \bar k} f^{\bar \jmath a} \bar f^{\bar k}_b f^b_l F^l \nn \\[1.5mm]
\a\a  \hspace{46pt} +\, h^{ab} \bar X_{a u} X_{b \bar \imath} + i h^{bc} h_{ab \bar \imath} \bar X_{c u} D^a\,, \\[1.5mm]
\a\a (m_0^2)_{uv} = - 2\, R_{u \bar s v \bar t} \bar X_A^{\bar s} L^A \bar X_B^{\bar t} \bar L^B
+ \s{\sqrt{2}}\, \Omega_{u s} \nabla_v X_a^s f_i^a F^i -\, h^{ab} \bar X_{a u} \bar X_{b v} + \Gamma_{uv}^k V_{{\rm S}k}  \,, \\[1mm]
\a\a (m_0^2)_{ij} = \s{\frac i2} \nabla_i C_{jkl} \, \big(2\,f_a^k \bar f^a_{\bar q} F^l \bar F^{\bar q}  + f^k_a f^l_b D^a D^b \big) 
+ \s{\sqrt{2}}i\, \Omega_{us} X_a^s C_{ijk} \bar f^{ka} F^u \nn \\[0.5mm]
\a\a \hspace{46pt} -\, h^{ab} \bar X_{a i} \bar X_{b j} + 2i h^{bc} h_{ab\t{(}i} \bar X_{cj\t{)}} D^a + \Gamma_{ij}^k V_{{\rm S}k} \,, \\[1.5mm]
\a\a (m_0^2)_{u i} = \s{\sqrt{2}}\, \Omega_{us} \big(\nabla_v X_a^s f_i^a F^v \!+ i X_a^s C_{ijk} \bar f^{j a} F^k\big) \!
- h^{ab} \bar X_{au} \bar X_{bi} + i\, h^{bc} h_{abi} \bar X_{cu} D^a .
\eea
The mass matrix of the fermions is instead
\bea
\a\a (m_{1/2})_{uv} = - \s{\sqrt{2}}\, \Omega_{\t{(}uw} \nabla_{v\t{)}} X_A^w L^A \,, \\[1mm]
\a\a (m_{1/2})_{ij} = - i C_{ijk} \, f_a^k \bar f^a_{\bar l} \bar F^{\bar l} \,, \\[1mm]
\a\a (m_{1/2})_{ab} =  - i C_{ijk} f^i_a f^j_b \, F^k \,, \\[0mm]
\a\a (m_{1/2})_{ia} = \s{\sqrt{2}}\, \bar X_{ai} - \s{\frac 1{\!\sqrt{2}}} C_{ijk} f^j_a f^k_b\, D^b \,, \\[-0.5mm]
\a\a (m_{1/2})_{ui} = - \s{\sqrt{2}}\, \Omega_{u v} X_a^v f_i^a \,, \\[1mm]
\a\a (m_{1/2})_{ua} = \s{\sqrt{2}} \bar X_{au} \,.
\eea
Finally, the mass matrix of the vectors is
\bea
\a\a (m_1^2)_{ab} = 2\, \big(X_{\t{(}a}^u \bar X_{b\t{)}u} + X_{\t{(}a}^i \bar X_{b\t{)}i} \big)\,.
\eea
A straightforward computation gives:
\bea
\a\a {\rm tr} [m_0^2] = 4\, \nabla_u X^v_A L^A \nabla^u \bar X_{Bv} \bar L^B 
+ 2\, R_{i \bar \jmath}\, \big(F^i \bar F^{\bar \jmath} \!+\! f_a^i \bar f^a_{\bar q} \bar f_b^{\bar \jmath} f^b_{p} F^p \bar F^{\bar q} 
\!+\! f^i_a \bar f^{\bar \jmath}_b D^a D^b \big) \nn \\[1mm]
\a\a \hspace{42pt} +\, 2\, h^{ab} \big(5\, \bar X_{au} X_b^u + \bar X_{ai} X_b^i \big) 
- 4 i h^{bc} h_{abi} X_c^i D^a + {\rm h.c.} \,,\\[1.5mm]
\a\a {\rm tr} [m_{1/2}^2] = 2\, \nabla_u X^v_A L^A \nabla^u \bar X_{Bv} \bar L^B 
+ R_{i \bar \jmath}\, \big(F^i \bar F^{\bar \jmath} \!+\! f_a^i \bar f^a_{\bar q} \bar f_b^{\bar \jmath} f^b_{p} F^p \bar F^{\bar q} 
\!+\! f^i_a \bar f^{\bar \jmath}_b D^a D^b \big) \nn \\[0.5mm]
\a\a \hspace{52pt} +\, 4\, h^{ab} \big(2\,\bar X_{au} X_b^u + \bar X_{ai} X_b^i \big)
- 2 i\, h^{ab} h_{bci} X_a^i D^c + {\rm h.c.} \,, \\[1mm]
\a\a {\rm tr} [m_1^2] = 2\, h^{ab} \big(\bar X_{au} X_b^u + \bar X_{ai} X_b^i \big) \,.
\eea
It follows that the supertrace of the mass matrix vanishes \cite{HKLR}:
\bea
{\rm str}[m^2] \b\equiv\b {\rm tr} [m_0^2] - 2\, {\rm tr} [m_{1/2}^2] + 3\, {\rm tr} [m_1^2] \nn \\[1mm]
\b=\b 0 \,.
\eea
This result also follows directly from (\ref{strcv}) and the properties that the Christoffel symbols are related to the 
derivative of the gauge kinetic function, the special form of the Ricci tensor and finally that the trace of the 
charge matrix satisfies the generalization $\nabla_u X_a^u = 0$ of (\ref{tracelessh}) in the hyper multiplet 
sector and (\ref{tracelessv}) in the vector multiplet sector.

\subsection{Metastability}

The possible vacua of the theory correspond to points in the scalar manifold that 
satisfy the stationarity conditions $V_{{\rm S}u} = V_{{\rm S}i} = 0$, which imply
\bea
\a\a - \s{\sqrt{2}}\, \Omega_{uw} \big(\nabla_{v} X_A^w L^A F^v\! + X^w_a f_i^a F^i\big) + i \bar X_{a u} D^a = 0 \,, \\
\a\a - \s{\frac i2} C_{ijk} \big(2\,f_a^j \bar f^a_{\bar q} F^k \bar F^{\bar q}\! + f^j_a f^k_b\,D^a D^b\big) 
- \s{\sqrt{2}}\, \Omega_{uw} X^w_a f_i^a F^u + i \bar X_{a i} D^a = 0
\label{statvecgen}
\eea
The relation (\ref{relDFF}) between the values of the $F^i,F^u$ and $D^a$ auxiliary fields, becomes
\bea
\a\a i \big(\nabla_u X_{a \bar v} F^u \bar F^{\bar v} \!+ 
(X_a^k h_{bck} f_i^b \bar f_{\bar \jmath}^c + f_{ac}^{\;\;\,b} f_i^c f_{\bar \jmath b})F^i \bar F^{\bar \jmath} \big) \nn \\
\a\a -\, \big(X_{\t{(}a}^u \bar X_{b\t{)}u} \!+\! X_{\t{(}a}^i \bar X_{b\t{)}i}\big) D^b + \s{\frac 12} f_{ab}^{\;\;\;d} k_{dc} \,D^b D^c = 0 \,.
\eea

On the vacuum, one has $\delta \chi^u = \s{\sqrt{2}} \epsilon F^u$, $\delta \psi^i = \s{\sqrt{2}} \epsilon F^i$,
$\delta \lambda^a = i \epsilon D^a$ and similarly $\hat \delta \chi^u = \s{\sqrt{2}} \hat \epsilon \hat F^u$, 
$\hat \delta \psi^i = \s{\sqrt{2}} \hat \epsilon \hat F^i$, $\hat \delta \lambda^a = i \hat \epsilon \hat D^a$,
and the first and second supersymmetries are spontaneously broken respectively if some of the auxiliary 
fields $F^u$, $F^i$, $D^a$ or some of the $\hat F^u$, $\hat F^i$, $\hat D^a$ are non-vanishing. 
The order parameters are the norms of the vectors defined by these two sets of quantities. Thanks to 
the identities $\hat F^u \hat {\bar F}_u = F^u \bar F_u + \frac 12 K^{a V} K_a^H$, 
$\hat F^i \hat {\bar F}_i = \frac 12 D^a D_a - \frac 12 K^{a V} K_a^H$ and 
$\frac 12 \hat D^a \hat D_a = F^i \bar F_i$, these two norms do actually as before coincide, 
and define in two equivalent ways related to the two supersymmetries the scalar potential energy:
$V_{\rm S} = F^u \bar F_u + F^i \bar F_i + \frac 12 D^a D_a 
= \hat F^u \hat {\bar F}_u + \hat F^i \hat {\bar F}_i + \frac 12 \hat D^a \hat D_a$.
In such a situation, there are then as usual two massless Goldstini, associated to the two 
independent supersymmetries and given by:
\be
\eta = \s{\sqrt{2}} \bar F_u \chi^u + \s{\sqrt{2}} \bar F_i \psi^i + i D_a \lambda^a \,,\;\;
\hat \eta = \s{\sqrt{2}} \hat {\bar F}_u \chi^u + \s{\sqrt{2}} \hat {\bar F}_i \psi^i + i \hat D_a \lambda^a \,,\;\;
\ee
With a bit of work one can verify that the stationarity conditions and the various identities related to 
gauge invariance imply that these are always flat directions of the fermion mass matrix:
\be
m_\eta = 0 \,,\;\; m_{\hat \eta} = 0 \,.
\label{resultmetagen}
\ee
In the situation under consideration, the two supersymmetries may again only be broken 
simultaneously.\footnote{See \cite{LST} for a recent systematic discussion on the conditions under which
one may have partial supersymmetry breaking. As explained in previous section, at stationary points 
with such a partial supersymmetry breaking, the two Goldstini must become degenerate and represent 
only one massless Goldstone fermion.}
As before, the sGoldstini are linear combinations of scalars and vectors, but what is 
relevant is their projection along the scalar field space. One then gets four 
independent real linear combinations, corresponding to the projection of the complex Goldstino 
vectors $\eta^\theta = (\s{\sqrt{2}} F^u,\s{\sqrt{2}} F^i)$ and $\hat \eta^\theta = (\s{\sqrt{2}} \hat F^u,\s{\sqrt{2}} \hat F^i)$:
\bea
\varphi_+ \b=\b \bar F_u q^u + \bar F_i \phi^i + F_{\bar u} \bar q^{\bar u} + F_{\bar \imath} \bar \phi^{\bar \imath}\,,\;\;
\varphi_- = i \bar F_u q^u + i \bar F_i \phi^i -i F_{\bar u} \bar q^{\bar u} - i F_{\bar \imath} \bar \phi^{\bar \imath} \,,\\
\hat \varphi_+ \b=\b \hat{\bar F}_u q^u + \hat{\bar F}_i \phi^i + \hat F_{\bar u} \bar q^{\bar u} + \hat F_{\bar \imath} \bar \phi^{\bar \imath}\,,\;\;
\hat \varphi_- = i \hat{\bar F}_u q^u + i \hat{\bar F}_i \phi^i -i \hat F_{\bar u} \bar q^{\bar u} - i \hat F_{\bar \imath} \bar \phi^{\bar \imath} \,.
\eea
The masses of these four scalar modes can now be computed by evaluating the scalar 
mass matrix along the directions $\varphi_+^\Theta = (F^u,F^i, \bar F^{\bar u},\bar F^{\bar \imath})$, 
$\varphi_-^\Theta = (i F^u, i F^i, - i \bar F^{\bar u},- i \bar F^{\bar \imath})$, 
$\hat \varphi_+^\Theta = (\hat F^u, \hat F^i, \hat {\bar F}^{\bar u}, \hat {\bar F}^{\bar \imath})$
and $\hat \varphi_-^\Theta = (i \hat F^u, i \hat F^i, - i \hat {\bar F}^{\bar u}, - i \hat {\bar F}^{\bar \imath})$, and dividing by the length of 
these vectors, which is $2 (F^u \bar F_u + F^i \bar F_i)$ for the first two and 
$2 (\hat F^u \hat {\bar F}_u + \hat F^i \hat {\bar F}_i)$ for the last two. 

One may at this point proceed in computing more explicitly the above sGoldstino masses and 
trying to simplify them as much as possible, in order to extract some information that has 
a simple-enough form to be useful. We will not attempt to do this here, but hope to 
examine this problem elsewhere, now that it has been set up in full detail within $N=2$ 
rigid supersymmetry. We again expect only one $SU(2)$-invariant information, generalizing 
those found for situations involving only hyper multiplets or only vector multiplets. 
It is however not entirely obvious how to proceed to extract such an information within the 
$N=1$ superspace formalism used in this paper, where the $SU(2)$ symmetry is not manifest. 
In particular, the way the four sGoldstini must be combined to yield this $SU(2)$-invariant information 
cannot be easily determined a priori, and as a matter of fact it looks different in the two 
subcases involving respectively only hyper or only vector multiplets. We believe that to clarify 
this issue it might be useful to compare with a manifestly $SU(2)$-covariant formalism, like for 
instance the on-shell approach of \cite{DFF,ABCDFFM}. 

\section{Conclusion}
\setcounter{equation}{0}

In this work, we have performed a general study of the conditions under which 
vacua breaking spontaneously supersymmetry may be at least metastable,
in the context of general $N=2$ non-linear sigma-models. To do so we have 
relied on a construction of these models based on $N=1$ superspace, which
allows to emphasize their peculiarities as special cases of $N=1$ non-linear 
sigma-models. We have then systematically applied to these models the strategy
of looking at the masses of the scalar modes belonging to the Goldstino 
would-be multiplet, which are the most dangerous modes for metastability.

We have been able to reproduce the two known no-go theorems available in the 
supergravity context, concerning theories with only hyper multiplets \cite{GLS} 
and only Abelian vector multiplets \cite{Many}. We have then clarified the origin 
of these sharp results, taking the perspective that such theories are particular
cases of $N=1$ theories involving only chiral multiplets, where supersymmetry breaking 
is controlled only by $F$ auxiliary fields. We have then studied in quite some detail 
the case of theories with only vector multiplets but with general non-Abelian gaugings,
giving evidence that no obstruction against achieving metastability subsists in this case.
From the $N=1$ perspective, these are special classes of theories involving
chiral multiplets in the adjoint representation and vector multiplets, where 
supersymmetry breaking is controlled not only by $F$ auxiliary fields but also by 
$D$ auxiliary fields. Finally, we have set up the study of general theories involving 
both hyper and vector multiplets, although we did not present any simple general 
result in this case. From the $N=1$ point of view, these are particular cases of 
theories involving chiral multiplets both in the adjoint representation and in 
more general representation, as well as vector multiplet, where the process 
of supersymmetry breaking is controlled both by $F$ and $D$ auxiliary fields.
We think that the effect of the latter should generically allow for metastable 
supersymmetry breaking vacua, since for general $N=1$ theories it is known
to systematically improve the situation compared to the effect of the former.

We believe that the results derived in this paper should be useful to address 
the general question of what are the mandatory ingredients to obtain metastable 
de Sitter vacua in $N=2$ supergravity theories. The results that we 
have obtained in the analysis of the corresponding problem in the rigid limit suggest that 
the only necessary ingredient is that from the $N=1$ perspective supersymmetry 
breaking should receive not only $F$-type but also $D$-type contributions. 
This requires either non-Abelian gauge groups, or charged hyper multiplets,
or both of these ingredients. 

Concerning the implications of the necessary conditions for metastable 
supersymmetry breaking for potentially realistic string models, one should 
keep in mind that these are described by $N=1$ effective theories, but with a 
hidden sector that displays many features of $N=2$ or even $N=4$ models. 
As a result, applying the $N=1$ constraints is too optimistic,
whereas applying $N=2$ or even $N=4$ constraints is too restrictive. One may
then try to consider the intermediate framework of $N=1$ theories obtained by 
truncations of $N=2$ or $N=4$ supersymmetries. In this kind of truncations, the projection
getting rid of the additional supersymmetries also eliminates the corresponding 
additional sGoldstini and the resulting implications on metastability. As a result,
one should get conditions that are stronger but have the same form as those for 
general $N=1$ theories. These should account for the possibility of starting 
from an unstable supersymmetry breaking vacuum and getting a metastable 
one by a truncation, where the tachyonic sGoldstini are projected out. 
For instance, it has been recently shown in \cite{RR} that the metastable $N=2$ 
de Sitter vacua of \cite{FTV} can be obtained by truncations of the unstable 
$N=4$ de Sitter vacua of \cite{RWP,RWPT}, which can themselves be related 
to truncations  of the unstable $N=8$ de Sitter vacua discussed in \cite{KLPS}.
It should be similarly possible to construct stable $N=1$ de Sitter vacua by 
truncating unstable $N=2$ de Sitter vacua. Since a detailed general description
of this kind of truncations is available \cite{ADF1,ADF2}, it would be interesting to 
perform a general study of the metastability conditions in this case.

During the completion of this work, the interesting paper \cite{AB} appeared, 
which explores the possibility of constructing a low-energy effective description of 
$N=2$ theories below the supersymmetry breaking scale in terms of constrained 
superfields, containing only the two Goldstini and no other light state. It was found 
that under the assumption of an $SU(2)_R$ symmetry, such an effective theory 
does not exist. This fact was interpreted as signaling the impossibility of 
achieving metastable supersymmetry breaking in such $N=2$ theories. 
This is compatible with what we found in this paper for theories involving only 
hyper multiplets or only vector multiplets without $N=2$ Fayet-Iliopoulos terms,
where some of the sGoldstini are unavoidably massless or tachyonic. 
We believe that the algebraic obstruction uncovered in \cite{AB} should 
correspond to the physical obstruction studied here against achieving a positive 
mass squared for all the sGoldstini, since whenever one of the sGoldstini is tachyonic 
one clearly cannot define a sensible low-energy effective theory for just the Goldstini. It would be very interesting to make this 
connection more precise and try to exploit it to study more efficiently the most general 
case of $N=2$ theories involving both hyper and vector multiplets as well as 
$SU(2)_R$-breaking Fayet-Iliopoulos terms.

\vskip 20pt

\noindent
{\bf \Large Acknowledgements}

\vskip 10pt

\noindent
This work was partly supported by the Swiss National Science Foundation.
We thank L.~Alvarez-Gaum\'e, I.~Antoniadis, M.~Buican, G.~Dall'Agata, S. Ferrara, 
M.~G\'omez-Reino and J.~Louis for useful discussions.

\end{document}